\documentclass[natbib, smallextended]{svjour3}     % onecolumn (second format)
\smartqed  % flush right qed marks, e.g. at end of proof
\usepackage{natbib}

\usepackage{graphicx}

\usepackage{amsmath}
\usepackage{amssymb}
\usepackage{color}
\begin{document}

\title{Modelling Quasi-Periodic Pulsations in Solar and Stellar Flares}

\titlerunning{Modelling Quasi-Periodic Pulsations in Solar and Stellar Flares}
%\titlerunning{Waves and pulsations in flares}

\author{J.~A. McLaughlin \and V.~M. Nakariakov \and M. Dominique \and P. Jel\'inek \and S. Takasao}

\authorrunning{J.~A. McLaughlin et al.}

%----------------------------------------------------------------------------------------------------------------------------------

\institute{
J. A. McLaughlin \at
Northumbria University, Newcastle upon Tyne, NE1 8ST, UK,\\
\email{james.a.mclaughlin@northumbria.ac.uk}
\and
V. M. Nakariakov \at
Centre for Fusion, Space and Astrophysics, University of Warwick, Coventry CV4 7AL, UK,\\
School of Space Research, Kyung Hee University, Yongin, 446-701, Gyeonggi, Korea,\\
St. Petersburg Branch, Special Astrophysical Observatory, Russian Academy of Sciences, 196140, St Petersburg, Russia, \email{v.nakariakov@warwick.ac.uk}
\and
M. Dominique \at
Royal Observatory of Belgium/STCE, 3, Avenue Circulaire, 1180 Brussels, Belgium, \email{marie.dominique@oma.be}
\and
P. Jel\'inek  \at
Institute of Physics and Biophysics, Faculty of Science, University of South Bohemia, CZ - 370 05, Branišovská 1760, \v{C}esk\'e Bud\v{e}jovice, Czech Republic, \email{pjelinek@prf.jcu.cz}
\and
S. Takasao \at
Department of Physics, Nagoya University, Nagoya, Aichi 464-8602, Japan,\\
\email{takasao@nagoya-u.jp}
}

%----------------------------------------------------------------------------------------------------------------------------------

\date{Received: date / Accepted: date}

\maketitle

\begin{abstract}
Solar flare emission is detected in all EM bands and variations in flux density of solar energetic particles. Often the EM radiation generated in solar and stellar flares shows a pronounced oscillatory pattern, with characteristic periods ranging from a fraction of a second to several minutes. These oscillations are referred to as quasi-periodic pulsations (QPPs), to emphasise that they often contain apparent amplitude and period modulation. We review the current understanding of quasi-periodic pulsations in solar and stellar flares. In particular, we focus on the possible physical mechanisms, with an emphasis on the underlying physics that generates the resultant range of periodicities. These physical mechanisms include MHD oscillations, self-oscillatory mechanisms, oscillatory reconnection/reconnection reversal, wave-driven reconnection,  two loop coalescence, MHD flow over-stability, the equivalent LCR-contour mechanism, and thermal-dynamical cycles. We also provide a histogram of all QPP events published in the literature at this time. The occurrence of QPPs puts additional constraints on the interpretation and understanding of the fundamental processes operating in flares, e.g. {{ magnetic energy liberation and  particle acceleration.}} Therefore, a full understanding of QPPs is essential in order to work towards an integrated model of solar and stellar flares.
\end{abstract}

%\PACS{52.35.Bj \and 94.30.cq \and 94.30.Ms \and 96.60.P- \and 96.60.Q- \and 96.60.qe}

\tableofcontents

\section{Introduction}\label{intro}

Flares constitute one of the most impressive manifestations of solar and stellar activity. They appear as a sudden increase of radiated flux, detectable in a broad range of wavelengths, going from gamma rays to radio. Since the first observations of a solar flare by \citet{1859MNRAS..20...13C} and \citet{1859MNRAS..20...15H}, countless detections were reported of flares on the Sun as well as on other stars (\citeauthor{1924BAN.....2...87H} is credited of the first stellar flare detection in 1924). However, flares have not yet revealed all their secrets. 

According to the most often invoked flare model, the CSHKP (\citealt{1964NASSP..50..451C}; \citealt{1968IAUS...35..471S}; \citealt{1974SoPh...34..323H}; \citealt{1976SoPh...50...85K}; \citealt{1992LNP...399....1S}) --- or \emph{standard} --- model, flares have their origin in magnetic reconnection that takes place in a coronal current sheet. The reconnection process accelerates particles in both the upwards and downward directions to non-thermal speeds. The latter, after propagating collisionlessly along magnetic field lines through the corona, eventually reach the denser chromosphere. There, they dissipate part of their energy by radiating (e.g. bremsstrahlung processes that produce hard X-rays) and by heating the ambient plasma, which results in the so-called chromospheric evaporation. This evaporated plasma, while cooling, will produce thermal emission essentially in the EUV and soft X-ray ranges. This model, although providing a detailed phenomenological description of most of the flare characteristics, keeps the main quantitative aspects elusive. In particular, the way the energy produced at the reconnection site is transported to the chromosphere remains highly debated. Obviously, the accelerated electron beam is a good energy propagation agent, but it is not clear whether it suffices to explain the huge amount of energy released during the flare process. Some authors suggested that part of the flare energy could rather be transported downward by Alfv\'en waves {{ (e.g. \citealt{2008ApJ...675.1645F}) or by thermal conduction  (e.g. \citealt{1978ApJ...220.1137A}; \citealt{1995ApJ...439.1034C}; \citealt{2006ApJ...642L.169M}).}}

The occurrence of waves and pulsations associated with flares puts additional constraints on the interpretation and understanding of the fundamental processes operating in both solar and stellar flares (e.g. particle acceleration, magnetic energy liberation). In this way, one can consider waves as both an integral part of flare dynamics as well as a potential diagnostic of the flare process. The overarching goal of solar and stellar flare modelling is thus to create an integrated plasma model which will, ultimately, create a coherent vision of reconnection, waves and particle acceleration processes in flares. This review paper considers one of these three key components: the modelling of waves and pulsations in solar and stellar flares. Specifically, we focus on quasi-periodic pulsations (QPPs) --- see \S\ref{sec:QPPs_section} --- but also briefly review other important wave processes in the appendices (\ref{Appendix : global_flare_waves} and \ref{Appendix : sunquakes}).

Note that this paper focuses on modelling QPPs, and their possible production by waves and pulsations, and thus is primarily a theoretical {\emph{modelling}} review. For a comprehensive {\emph{observational}} overview of solar flares see, e.g. \citet{2011SSRv..159...19F}. Detailed reviews of observational and forward modelling aspects of QPPs are summarised in {{\citet{2009SSRv..149..119N}; \citet{2010PPCF...52l4009N}; \citet{2016SoPh..291.3143V}, }} while the theoretical aspect, mainly the mechanisms based on standing MHD oscillations, is covered there too. 
In this paper we address  the QPP mechanisms developed in recent years, such as periodic reconnection, the magnetic tuning fork, and self-oscillatory processes, as well as some well-known mechanisms, e.g. the equivalent LCR contour and dispersive wave trains, which recently obtained observational support, but have not obtained sufficient attention in  previous reviews.

%\citet{2009SSRv..149..119N, 2010PPCF...52l4009N, 2016SoPh..291.3143V}

\subsection{Oscillations, self-oscillations, waves and pulsations}\label{defs}

Let us start with some terminology and definitions. According to the common knowledge, an {\emph{oscillation}} is any motion, effect or change of state that varies periodically between two values, i.e. there is a repetitive nature. However, clearly, this definition does not include a number of {{constraints}}, such as the finite duration of the oscillatory pattern, possible amplitude modulation, e.g., the decay, and frequency drifts. In the {{Fourier spectral domain}}, an oscillation is usually associated with a statistically significant peak, or a group of peaks in the case of an anharmonic pattern. But, again, this approach does not take into account the oscillation life time and the modulations. Thus, it is difficult to produce a mathematically rigorous definition of an oscillation in real data. {{In flaring signals}}, this difficulty is magnified by the intrinsic localisation of the quasi-oscillatory patterns in a certain time interval that is determined not only by the properties of the oscillation itself, but also by the duration of the emission in the flare. For example, in the gyrosynchrotron emission an oscillatory pattern is seen only during the operation of this mechanism, i.e. when there are non-thermal electrons in the oscillating plasma. Thus, we usually intuitively consider a {\emph{quasi-periodic pulsation}} (QPP) to be a quasi-repetitive pattern in the signal, which has at least three or four iterations --- the QPP cycles.

It is easier to define an oscillation in theoretical modelling. From this point of view, an oscillation is a quasi-periodic variation of certain physical parameters in the vicinity of a certain equilibrium. For example, it is the (quasi)-periodic dynamics of a load of the pendulum, or, in the case of solar flares, a (quasi)-periodic variation of the plasma density with respect to the equilibrium in a flaring loop. It should be pointed out that the equilibrium itself may vary during the oscillation, for example the equilibrium value of the density in the loop may change because of the ongoing chromospheric evaporation, or gradual variation of the loop length or width. Parameters of an oscillation, such as the amplitude and phase, are determined by the initial excitation. In general, in an oscillation there is a (quasi)-periodic {{transformation of the kinetic, potential, magnetic and thermal energy into each other. There is also the continuous sinking of the oscillation energy to the internal energy}}, and possibly radiation of the energy outward the oscillating system. Thus, an oscillation can be considered as a (quasi)-periodic competition between an effective restoring force and inertia. The oscillation period is determined by the properties of the oscillating system, an oscillator. In a certain time interval, oscillations may be driven by an external time-dependent force, resupplying the oscillation with energy. In this case the response of an oscillator to the external force consists of a combination of the natural oscillation and the driven oscillation. When the frequencies of the natural and driven oscillations are close to each other, the phenomenon of resonance occurs. 

An important class of oscillatory motions in dissipative and active\footnote{The medium could be considered as active if certain perturbations provoke the medium to release energy} media are self-sustained oscillations, also called {\emph{self-oscillations}}, auto-oscillations or oscillatory dissipative structures. Self-oscillations occur in essentially non-conservative systems  because of the competition between the energy supply and losses. In particular, in electronics self-oscillations are associated with the process of the conversion of the direct current in the alternate current of a certain frequency. Self-oscillatory motions are common in a number of dynamical systems, and the well-known examples are various musical instruments, radio-frequency generators, the heart, the clock \citep[see][for a comprehensive review]{2013PhR...525..167J}. Usually, the self-oscillation period depends on the amplitude. Despite the presence of dissipative and/or radiative losses, a self-oscillation may be decayless, because of the continuous extraction of the energy from the medium. This behaviour should not be confused with the driven oscillations mentioned above, as in the case of self-oscillations this energy supply comes from an essentially non-periodic source, e.g. the DC battery in a watch, or the steady wind causing the periodic shedding of aerodynamic vortices. In solar flares, a steady inflow of the magnetic flux towards the reconnection site could result in repetitive magnetic reconnection \citep[\lq\lq magnetic dripping\rq\rq,][]{2010PPCF...52l4009N} that should be considered as a self-oscillatory process. 

In contrast with regular oscillations, properties of self-oscillations, such as the period, shape of the signal, and amplitude are uniquely determined by the parameters of the system they are supported by, and are independent of the initial conditions. It makes them an excellent tool for seismological probing of the media and physical processes operating there. Hence, the search for and identification of self-oscillatory processes in solar and stellar impulsive energy releases is an interesting research avenue.

A {\emph{wave}} is a perturbation that propagates through space and time, which is usually accompanied by energy transference. Despite the common knowledge that a wave should be \lq\lq wavy\rq\rq, it is not necessary for the wave signal to be periodic. The main property of a wave is its propagation that is characterised by its phase and group speeds. More rigorously, a wave is a signal that, in the simplest, one-dimensional case, is described by the general solution to the wave equation, $f(z-Ct)$, where $z$ and $t$ are the spatial coordinate and time, and $C$ is the phase speed of the propagation. The function $f$ that describes the wave shape is an arbitrary, sufficiently smooth function that is determined by the excitation. In particular, it may be periodic, e.g. harmonic, or aperiodic, e.g. Gaussian. 

In a more general case the function $f$ can also gradually vary in time and space, as it is, e.g. in the presence of dissipation, mode conversion, or non-plane effects. If nonlinear effects are important, the speed $C$ may become a function of the amplitude, and the wave evolution is described by a certain evolutionary equation, e.g. the Burgers equation for magnetoacoustic waves, or the Cohen--Kulsrud equation for Alfv\'en waves. Nonlinear evolution of a wave usually leads to the deformation of the wave shape, e.g. the formation of the characteristic saw-tooth pattern in the case of nonlinear magnetoacoustic waves. Shock waves are a specific class of nonlinear wave motions, with the functions $f$ having an infinite gradient. 

In dispersive media or systems, signals with different frequencies have different phase and group speeds, for example a fast magnetoacoustic wave propagating within a system with a field-aligned inhomogeneity. In this case, different spectral components that are the results of the Fourier decomposition of the function $f$ propagate at different speeds, and an initially broadband signals evolves into a locally harmonic signal. This situation occurs, in particular, in the case of the waves on the surface of water, which leads to our everyday experience that a wave should be \lq\lq wavy\rq\rq. 

Similarly to self-oscillations, there could be \lq\lq self-waves\rq\rq, more often called \textit{autowaves} that appear in active media, when the passage of the wave causes the energy release that reinforces the wave. An example  of an autowave is the wave of flame. The speed, amplitude and other parameters of autowaves are determined by the properties of the medium. In solar physics, autowaves could occur, for example, as the \lq\lq wave\rq\rq of sympathetic flares: an energy release in the first flare ignites the next one that, in turn, ignites the third, etc., i.e. a \lq\lq domino effect\rq\rq. The progression of the quasi-periodic energy release site along the neutral line in a two-ribbon flare could also be produced by an autowave.

Waves and oscillations are closely related to each other. A standing wave that is a linear superposition of two oppositely propagating waves of the same amplitude, is usually called an oscillation in solar physics. Examples of these oscillations are the fundamental magnetoacoustic harmonics of coronal loops, such as kink, sausage, fluting, torsional and acoustic modes \citep[see, e.g.][for comprehensive reviews]{2012RSPTA.370.3193D, 2016SSRv..200...75N}. 

\subsection{Waves and pulsations generated by flares}\label{sec:waves generated by flares}

The dramatic energy release in flares can generate waves and pulsations in the elastic and compressive solar atmosphere. Firstly, there is the impulsive energy release of the flare itself; this can act as an impulsive driver for waves and pulsations. Additionally, during the huge magnetic {{restructuring}}  that accompanies the reconnection, the magnetic field below the reconnection site is believed to collapse in an \lq\lq{implosion}\rq\rq{} process (\citealt{2000ApJ...531L..75H}) that would very likely trigger waves too. Simply put, the flare is converting stored energy into various forms which we observe both directly and indirectly, and waves/pulsations/outflows are part of that energy conversion process (see, e.g. \S3.3 of \citealt{2011SSRv..158....5H}). 

Thus, there is a rich tapestry of wave-related phenomena associated with solar (and stellar) flares. This review focuses on QPPs, but other types of waves and pulsations associated with solar and stellar flares are discussed briefly in the appendices, where Appendix \ref{Appendix : global_flare_waves} considers global waves generated by {{CMEs and}} flares (including shock waves, blast waves, EIT waves, Moreton waves and \lq{flare waves}\rq) and Appendix \ref{Appendix : sunquakes}  considers sunquakes (another wave-like global phenomenon associated with flares).

\subsection{Quasi-periodic pulsations}\label{sec:QPPs_section}

Quasi-repetitive patterns have been detected in a variety of signals generated by flares. These are referred to as {\emph{quasi-periodic pulsations}} (QPPs), and  have been observed in radio, optical and X-ray emission of solar flares {{(e.g. \citealt{1983ApJ...271..376K}; \citealt{1983ApJ...273..783K}; \citealt{1985SoPh..100..465D}; \citealt{2001ApJ...562L.103A}; }}  \citealt{2008A&A...487.1147I}; \citealt{2009A&A...493..259I}; \citealt{2009SSRv..149..119N}; \citealt{2016ApJ...827L..30H}; \citealt{2016SoPh..291.3143V}; \citealt{2016ApJ...832...65Z}) and stellar flares (e.g. \citealt{2003A&A...403.1101M}; \citealt{2006A&A...456..323M}; \citealt{2005A&A...436.1041M}). These are not, rigorously speaking, oscillations or waves (see \S\ref{defs} for terminology), rather they are oscillation trains (short bursts of oscillations) or, in some cases, modulated oscillations, i.e. time-varying (in amplitude or period) oscillations.

\begin{figure*}
\centering
\includegraphics[scale=0.7]{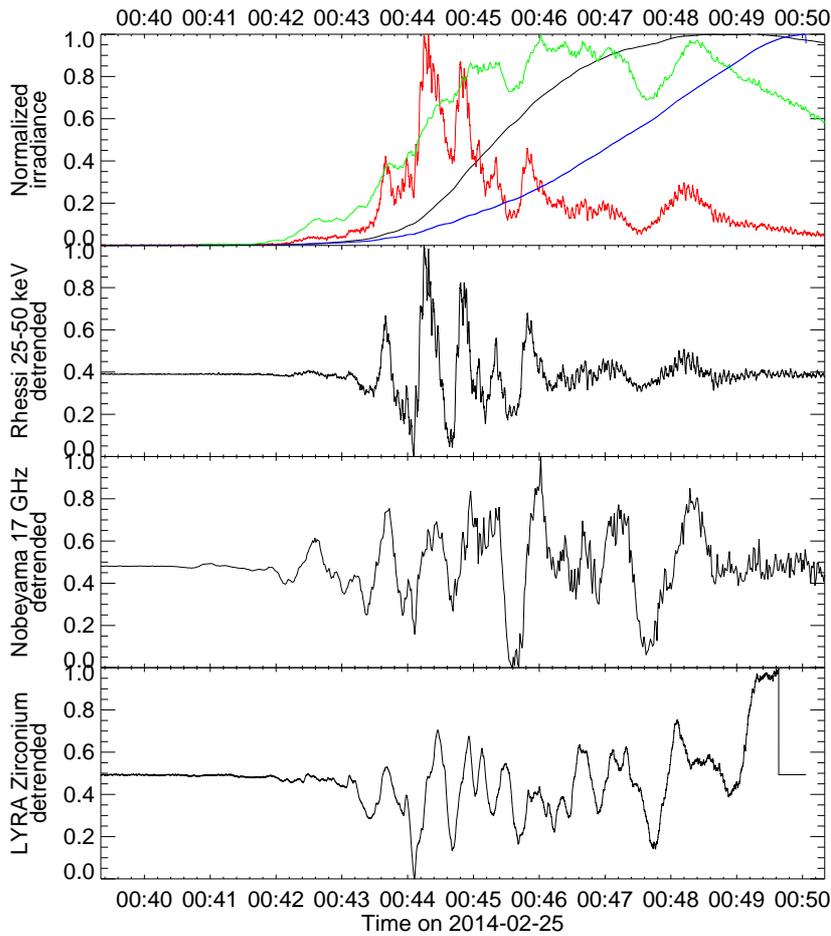}
\caption{Example of QPPs for the {X4.9 flare of 25 February 2014}. Top panel: the normalised flare time series from GOES 0.1--0.8~nm (black), Lyra 1--20~nm (blue), RHESSI 50--100~keV (red) and Nobeyama 17~GHz. Second, third and fourth panels: the same time series for respectively RHESSI, Lyra, and Nobeyama, detrended with a 50 s window. The QPPs appear to be remarkably synchronous. The data gap in the Lyra time series from 00:50~UT onwards is caused by a spacecraft manoeuvre.}
\label{QPP_example}      
\end{figure*}

An example of QPPs is illustrated in Figure \ref{QPP_example} for the X4.9 flare of {25 February 2014}. QPP oscillations with a period of $\sim$35~s are clearly visible as an oscillatory train in all displayed time series that cover the radio (Nobeyama Radio Polarimeters 17~GHz), the EUV (PROBA-2/Lyra 1--20~nm, see \citealt{2013SoPh..286...21D}) and the HXR ranges (RHESSI 50--100~keV, see \citealt{2002SoPh..210....3L}). Despite the very different ranges of energy considered, the oscillations are remarkably synchronous. 

In the top panel of Figure \ref{QPP_example}, the  green and red curves show the clear oscillatory pattern that is often displayed in the flare non-thermal emission. At the end of the 1960s, those oscillations were known to correlate well in the X-ray and radio bands, and a possible wave-origin had already been invoked \citep{1969ApJ...155L.117P}. Since then, numerous observations of these QPPs have been reported during solar flares, not only in non-thermal (see e.g. \citealt{1983ApJ...271..376K}; \citealt{2008A&A...487.1147I}), but also in thermal emission, with example cases in the visible (e.g. \citealt{1998SoPh..181..113J}; \citealt{2005ApJ...620.1101M}),  in the soft X-rays/EUV (e.g. \citealt{2012ApJ...749L..16D}; {{\citealt{2015ApJ...810...45B}}}) and in the ultraviolet ranges (e.g. \citealt{2016ApJ...823L..16T}), {{as well as simultaneously in both thermal and non-thermal emission (e.g. \citealt{2016ApJ...830..101B})}}. Such a global wavelength coverage tends to indicate that QPPs affect all layers of the solar atmosphere from the chromosphere to the corona.

\begin{figure*}
\begin{center}
\includegraphics[scale=0.3]{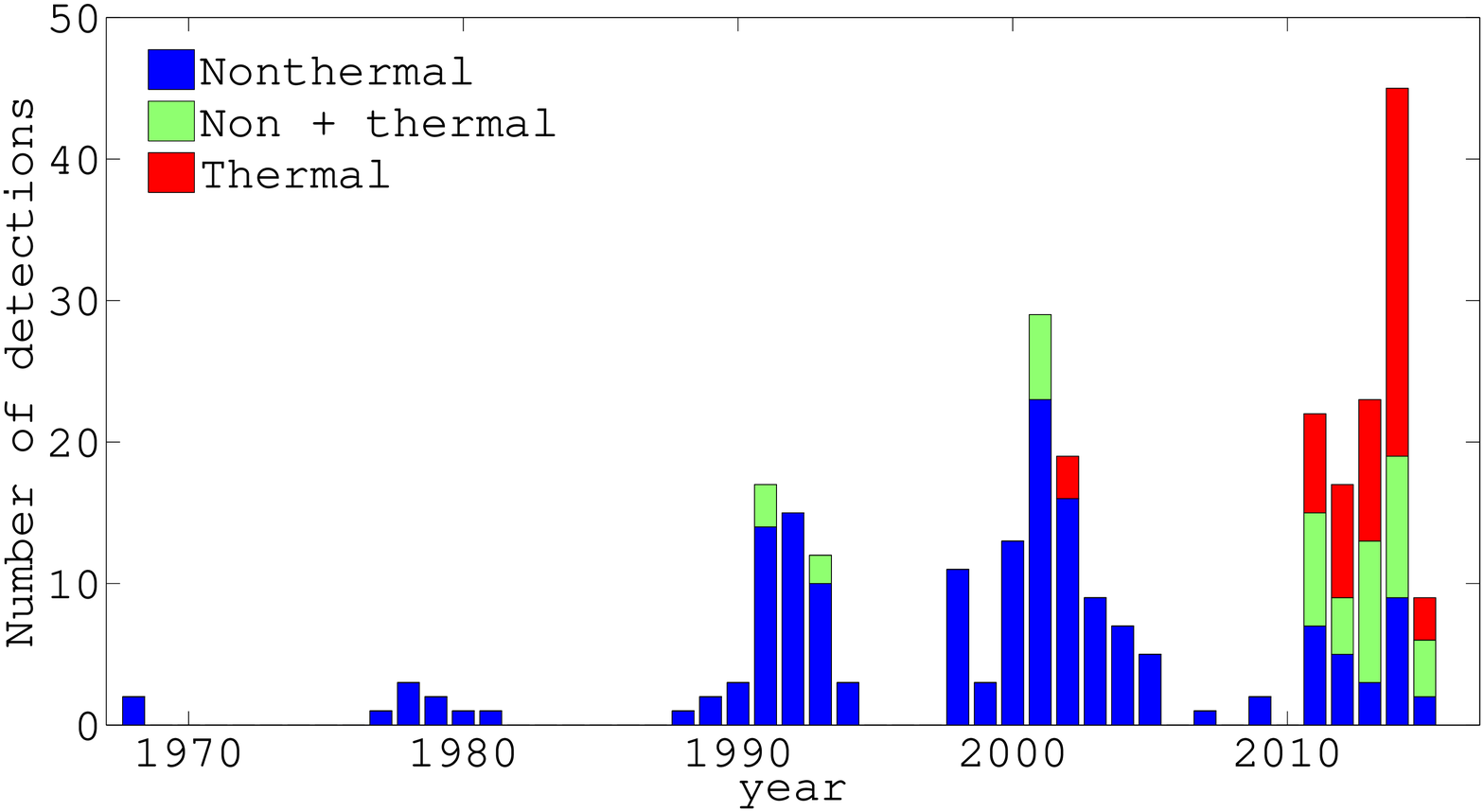} 
\includegraphics[scale=0.3]{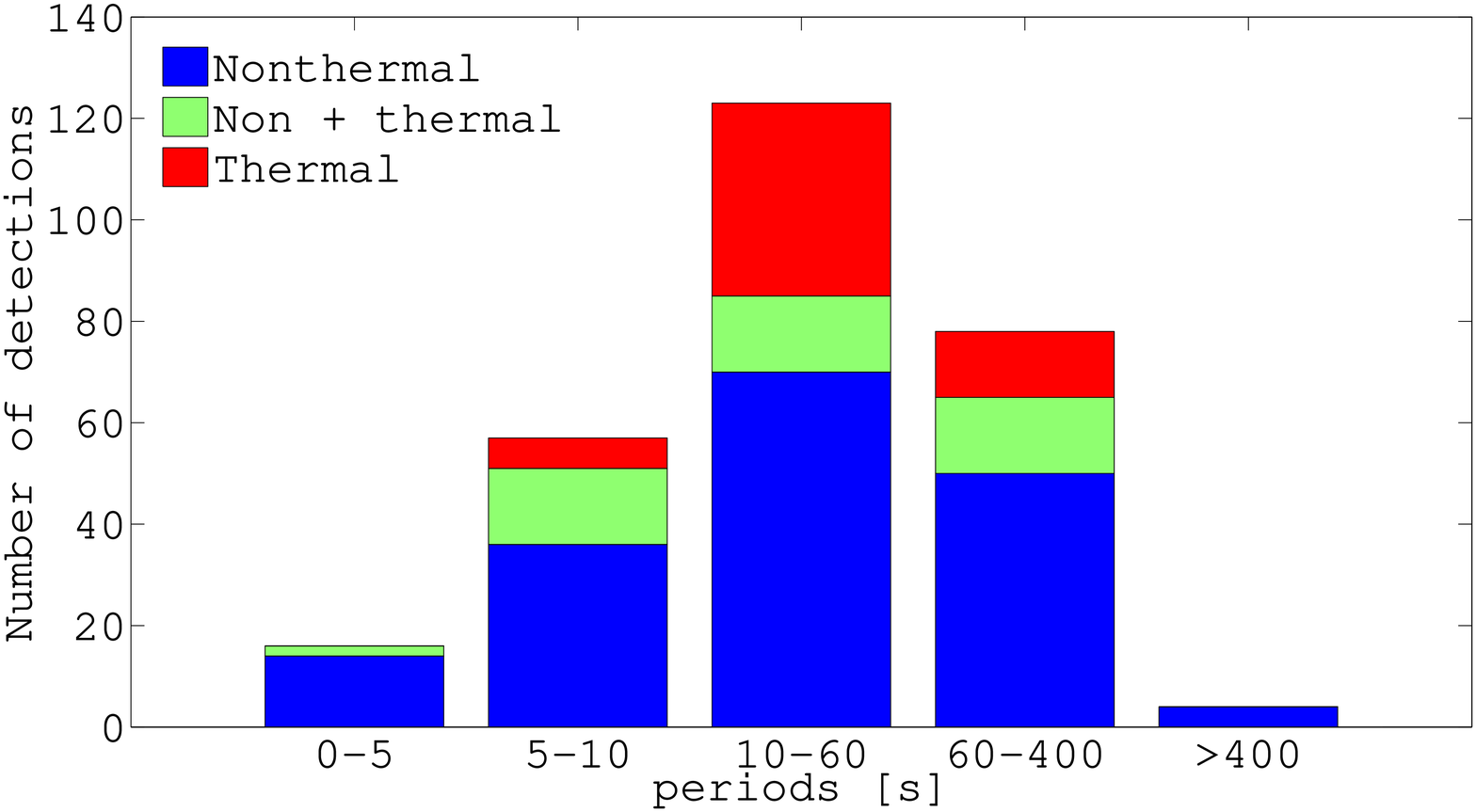}
\caption{
Properties of 278 QPP events reported in the literature. The top panel shows the time distribution, with the bin size being the calendar year. The bottom panel shows the distribution of detected periods. The blue colour shows QPPs detected in the non-thermal emission, red in thermal, and green detected in both thermal and non-thermal emissions simultaneously.
}
\label{figure_hist}
\end{center}
\end{figure*}

The web-page{\footnote{\textsl{http://www2.warwick.ac.uk/fac/sci/physics/research/cfsa/people/valery/research/qpp/}}} presents a catalogue that contains information about QPPs in solar flares, detected in various bands and with various instruments. The catalogue is based on the information provided in already published papers by various authors, is continuously updated, and at the moment contains  278 QPP events reported in the literature. Figure~\ref{figure_hist} illustrates the distribution of the detected QPPs in time and by the periods. In the cases of drifting periods we took the mean value of the period. We attribute an event to a QPP in the thermal emission if it was detected in {{EUV and/or soft X-rays}}, while QPPs in {{radio, microwave, {{visible light and white light (see below),}} hard X-ray and gamma-ray bands}} are considered as QPPs in the non-thermal emission. This separation is rather artificial, but may be useful for the choice of appropriate instrumentation for further studies of this phenomenon.

%We attribute an event to a QPP in the thermal emission if it was detected in EUV and/or soft X-rays, while QPPs in radio, microwave, visible light and white light (see below), hard X-ray and gamma-ray bands are considered as QPPs in the non-thermal emission. This separation is rather artificial, but may be useful for the choice of appropriate instrumentation for further studies of this phenomenon.

{{
We note that {\it{white light}} emission in flares is associated with non-thermal particles, whereas {\it{visible light}} includes various lines that could be more sensitive to thermal effects (for example, H{$\alpha$} is dependent mainly on temperature, not directly on the non-thermal process). Thus, we have attributed QPPs detected in visible light as non-thermal emission, but we emphasise that certain types of visible light could be classed as either thermal or non-thermal emission (as stated, the separation is rather artificial, but potentially useful). The classification is clearer for white light: with regards to white light and hard X-ray light curves, it has been observed that both behave in a similar manner in many flares (e.g. \citealt{1992PASJ...44L..77H}) and that both of these emissions in the impulsive phase of flares are caused by non-thermal electrons  (e.g. \citealt{2007ApJ...656.1187F}; \citealt{2010ApJ...715..651W}).}}

%{{Note here we have attributed QPPs detected in visible light as non-thermal emission. Here, {\it{white light}} emission in flares is associated with non-thermal particles, whereas {\it{visible light}} includes various lines that could be more sensitive to thermal effects (for example, H$_\alpha$ is dependent mainly on temperature, not directly on the non-thermal process). With regards to white light and hard X-ray light curves, it has been observed that both behave in a similar manner in many flares (e.g. \citealt{1992PASJ...44L..77H}) and that both of these emissions in the impulsive phase of flares are caused by non-thermal electrons (e.g. \citealt{2007ApJ...656.1187F}; \citealt{2010ApJ...715..651W}).}}

%H{$\alpha$}

The statistics of the QPP detections clearly correlates with the solar cycle, which is not a surprise, as the frequency of flares depends on the phase of the cycle. The recent increase in the detection of QPPs in the thermal emission is explained in the availability of EUV and soft X-ray instruments. The increase in the number of QPPs detected simultaneously in both thermal and non-thermal emission reflects also the growing interest in multi-instrumental studies of the QPP phenomenon, necessary for the exclusion of instrumental artefacts. The distribution of the detected periodicities is partly affected by the time resolution of the available instruments. These statistics confirms that QPPs are a rather common phenomenon that is intensively studied observationally. Detected periods range from a fraction of a second to several minutes, which means QPPs are detectable with the majority of modern solar instruments.

Stellar flares, far more energetic than typical solar flares have been observed on solar-like stars \citep{2012Natur.485..478M} leading to predictions of \lq{superflares}\rq. QPPs have been  reported in stellar flares throughout the whole spectrum (see e.g. \citealt{2016MNRAS.459.3659P} and references therein). Obviously, QPPs are neither a rare phenomenon, nor one that is limited only to the Sun. \citet{2016NatCo...711058K} analysed 48 superflare stars using the LAMOST telescope \citep{2012RAA....12.1197C} and suggested that solar flares and superflares most likely share the same underlying mechanism\footnote{{{Note that stellar QPPs are not to be confused with  Quasi-Periodic Oscillations (QPOs) which is a term used in the astrophysical literature in relation to  X-ray binaries (e.g. \citealt{1988SSRv...46..273L}; \citealt{1998ApJ...492L..59S}) and black hole binaries (e.g. \citealt{2012MNRAS.426.1701B}).}}}.

Furthermore, there has been a wealth of QPP detections in stellar flares using NASA's Kepler mission (\citealt{2010Sci...327..977B}), e.g. \citet{2014ApJ...797..122D} investigated the temporal morphology of white-light flares in Kepler data (\citealt{2016ApJ...829...23D} complied a Kepler catalogue of stellar flares). \citet{2013ApJ...773..156A} analysed the signal in the decay phase of the U-band light curve of a stellar megaflare and reported that the  oscillation was well approximated by an exponentially-decaying harmonic function. \citet{2015MNRAS.450..956B} analysed data from  257 flares in 75 stars to search for QPPs in the flare decay branch. \citet{2015ApJ...813L...5P} presented an analysis of a white-light stellar superflare observed by Kepler and detected a  multi-period QPP pattern.  \citet{2016MNRAS.459.3659P} studied QPPs  in the decay phase of white-light stellar flares and looked for correlations between QPP periods  and parameters of the host star. For the 56 flares with QPP signatures detected, no correlation was found between the QPP period and the stellar temperature, radius, rotation period or surface gravity, suggesting that QPPs are independent of global stellar parameters and are likely to be determined by the local parameters, e.g. of the flaring active region.

Systematic statistical studies of solar QPPs have also been performed. \citet{2010SoPh..267..329K} analysed twelve \lq{single-loop}\rq{} flares observed in the microwave band (i.e. in the non-thermal emission) and found statistically significant QPPs in ten of them. \citet{2015SoPh..290.3625S} found that  80\% of X-class flares from Cycle 24 {{(so far)}} display QPPs in thermal emission. More recently, \citet{2017A&A...597L...4L} reported on QPPs with periods that change depending on whether the pulsations have thermal or non-thermal components.

%\citet{2013ApJ...777..152S} showed that most of the solar flares they analysed exhibited  QPPs, usually during their impulsive and decay phases, making them an integral part of the flaring process.

{{While  \citet{2015ApJ...798..108I} claimed that QPPs are not statistically rigorous oscillations,  \citet{2016ApJ...833..284I} performed}} a large-scale search for evidence of signals consistent with QPPs in solar flares, focusing on the 1--300 second timescale, and concluded that 30\% of thermal events (GOES) and 8\% of non-thermal events (Fermi/GBM) show strong signatures consistent with the classical interpretation of a QPP, based on the significance level of the corresponding peak in the Fourier power spectrum. These estimations are rather conservative, as they address the search for stationary periodicities in the spectrum, while QPPs are often non-stationary, wavelet-like signals. There is a clear need for a definition of a QPP, which would account for the effects of coloured noises, regular trend and the intrinsic non-stationary nature of the quasi-oscillatory patterns in flaring light curves. 

QPP observations cover a wide range of periodicities (see Figure~\ref{figure_hist}). If sub-second periodicities are usually attributed to cyclic behaviours of self-organising systems driven by wave-wave or wave-particle interactions (see the reviews by \citealt{1987SoPh..111..113A}; \citealt{2008PhyU...51.1123Z}, as well as \citealt{1998A&A...334..314C} for a specific example), QPPs with periodicities from a few seconds to several minutes have often been attributed to MHD waves. Fast sausage modes are usually considered here, especially when dealing with sub-minute QPPs (e.g. \citealt{2003A&A...412L...7N}; \citealt{2005A&A...439..727M}), although slow magnetoacoustic (\citealt{2011ApJ...740...90V}; \citealt{2012ApJ...755..113S}) and fast kink modes (\citealt{2005A&A...440L..59F}) have been sometimes invoked to explain longer periodicities. 

However, MHD waves are not the only possible explanation for QPPs (see \citealt{2009SSRv..149..119N} for a discussion). The initial electron acceleration process, if being itself quasi-periodic, would also result in a spectrally broad modulation of the observed flux, both thermal and non-thermal. This mechanism was for example invoked by \citet{1983ApJ...271..376K} to explain the well-known Seven-Sisters Flare of 7 June 1980.

\section{Physical mechanisms underpinning QPP generation}\label{physmechqpp}

The motivation to understand the physical mechanism(s) responsible for QPPs  is clear:  the frequent occurrence of QPPs in flaring light curves puts additional constraints on the interpretation and understanding of fundamental flare physics. Thus, the rest of this review focuses on the discussion of the physical mechanisms proposed for the generation of QPPs in solar and stellar flares.

Whether QPPs are caused by MHD waves or an alternative mechanism(s) is a highly debated question and might depend on the considered range of periodicities. This section summarises the state-of-the-art understanding of each of those processes and aims to pin-point the spectral and temporal characteristics of the QPPs that each of them would produce, so as to help diagnosing the origin of QPPs in the various observational cases. 

%\textbf{Physical mechanisms / potential explanations (including identification mechanisms/non-ambiguous predictions)
%%When describing/detailing each mechanism, I think it would be good to state clearly:
%\begin{itemize}
%\item{What range of periods the mechanism can create/estimate/address.}
%\item{What underlying physics determines these periodicities}
%\item{How much this mechanism can be proven/identified in observations.}
%\item{What theory/advances should be done in the modelling as a next step, e.g. parametric studies.}
%\end{itemize}
%}

\subsection{MHD oscillations}\label{sec:coronal_seismology}

{{Some of the}} observed periods of QPPs coincide {{with}} the order of magnitude {{of}} MHD waves and oscillations that are abundantly detected in the corona (and well resolved both spatially and temporally).
 
\begin{figure*}
\begin{center}
\includegraphics[width=3.0in]{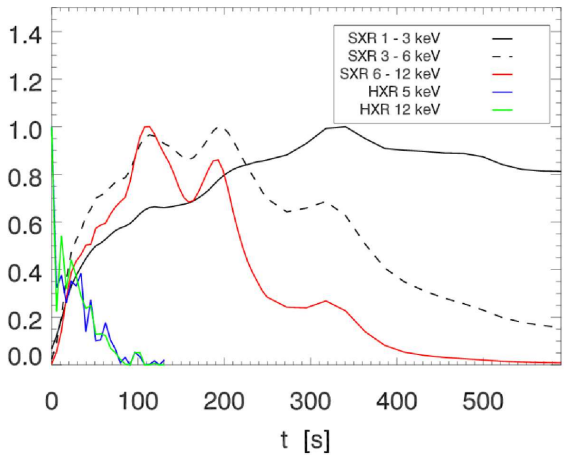}
\caption{
Light curves of the soft X-ray emission (black and red lines) and hard X-ray emission (blue and green lines) at different photon energy bands (see the inset legend). A few post-flare oscillations are visible in the soft X-ray light curves. From \citet{2016A&A...585A.159P}; their Figure 12, model V.
}
\label{figure_PINTO}
\end{center}
\end{figure*}

Coronal plasma flows or rearrangement of magnetic fields can cause the displacement of  coronal loops, filaments and streamers, which can result, for example, in transverse oscillations of these coronal structures. The initial energy deposition must come from somewhere, and the dramatic energy deposition from a flare could be the origin of such a driver (there are other potential origins, of course). In this sense, the flare is invoked as an impulsive perturbation, and that  impulsive energy release could be modelled as a thermal pressure pulse as well as a magnetic, velocity and/or heat  perturbation to the system. Such perturbations can be {\it{external}} to a loop system  (e.g. \citealt{2002ApJ...574..440O}; \citealt{2008ApJ...682.1338M}) or {\it{internal}} to a loop system. For example, in the latter case, \citet{2004A&A...414L..25N} studied the evolution of a coronal loop in response to an impulsive energy release and found that the evolution of the loop density exhibits quasi-periodic oscillations associated with the second standing harmonics of an acoustic wave (note that the slow magnetoacoustic oscillations --- since their study was limited to 1D --- could also be interpreted as the second, standing slow magnetoacoustic mode of the loop). Here, the perturbation was modelled as the response of the loop to a flare-like impulsive heat deposition at a chosen location.  \citet{2004A&A...422..351T} extending this work to look at  how the locations of the heat deposition  affects the mode excitation; it was found that excitation of such oscillations is independent of the heat-deposition location within the loop. On the other hand, numerical simulations of the response of a coronal loop to an impulsive heat deposition at one chromospheric footpoints demonstrates the effective excitation of the fundamental acoustic mode \citep{2005A&A...438..713T}. \citet{2016A&A...585A.159P} developed a model of the thermal and non-thermal emission produced during the evolution of kink-unstable twisted coronal loops in a flare. Their modelling showed post-flare oscillations, which could be interpreted as QPPs, in the soft X-ray emission, see Figure \ref{figure_PINTO}. {{\citet{2016ApJ...830..110C} investigated QPPs observed in the decay phase of solar and stellar flares in X-rays, and proposed that the underlying mechanism responsible for the stellar QPPs is a natural MHD  oscillation in the flaring or adjacent coronal loops.}}

In this sense, the flare is invoked as a justification for a source region and energy provider, but that once the finite-duration internal/external excitation occurs, we will get free MHD oscillations of the emitting plasma. Thus, we are now within the field of  {\emph{coronal seismology}} and so the observed parameters  tell us diagnostic information  about the medium and the oscillating structure itself (e.g. the magnetic field strength of an oscillating coronal loop; information about the heating function, transport coefficients, and fine sub-resolution structuring) rather than about the driver (be that a flare or other). In other words, the period will be independent of the flare energy and so coronal seismology tells us about the local conditions in flaring active regions, rather than the flare itself. Coronal seismology is a significant field in its own right and readers are {{recommended}} to consult the comprehensive reviews in this area \citep[e.g. see][and references therein]{2005RSPTA.363.2743D, 2012RSPTA.370.3193D}.

%The flare-driven and flare-excited waves mentioned in this section are waves/oscillations; there are variations about an equilibrium state. However,  the flare is invoked as an impulsive forcing (or excitation) term and the resultant periodic behaviour is due to the MHD modes of oscillation of the coronal structures. Thus,  the periodicity comes from the  global parameters of the oscillating loop not from the (flare) driver.

\subsection{QPPs periodically triggered by external waves}\label{periodic triggering}

%The flare-driven and flare-excited waves mentioned in this section are waves/oscillations; there are variations about an equilibrium state. However,  the flare is invoked as an impulsive forcing (or excitation) term and the resultant periodic behaviour is due to the MHD modes of oscillation of the coronal structures. Thus,  the periodicity comes from the  global parameters of the oscillating loop not from the (flare) driver.

{{In the previous subsection, the flare is invoked as an impulsive forcing term, hence the periodicity comes from the global parameters of the oscillating loop, not from the driver. However, in some cases MHD waves and oscillations may affect, or back-react on, the flaring process.}} Let us consider the trigger mechanism for flares. The pre-flare stage, i.e. before the primary energy release  (the impulsive phase), is one of energy storage. By definition, flares are the rapid release of energy stored previously in the magnetic field, and the total flare energy {{is of the same order of magnitude as  the amount of magnetic free energy, while the specific fraction is still debated in the literature (e.g. \citealt{2012ApJ...759...71E} reports  that for large  solar eruptive events, approximately 30\% of  the available  non-potential  magnetic energy is released).}} Reconnection must be at the heart of this  energy release and thus waves must play a role here, namely that it is known that steady-state reconnection models generate not only outflows/waves but also {\emph{require}} inflows/waves (e.g. \citealt{1957JGR....62..509P}; \citealt{1958IAUS....6..123S}; \citealt{1964NASSP..50..425P}). This is for example the case in the  CSHKP standard flare model which has a {\emph{null point}} --- a location where the magnetic field, and hence the Alfv\'en speed, is zero, at least in a certain plane --- as part of its magnetic topology.

In order to model and understand the  pre-flare stage, one must understand how a stable magnetic configuration (which has sufficient stored magnetic free energy) becomes unstable in such a way as to produce a rapid and dramatic  energy release. There are many models of how a magnetic topology can store magnetic energy \citep[e.g.][]{2013SoPh..288..481R, 2015ApJ...806....9K} or emerge  with sufficient magnetic free energy \citep[e.g.][]{1977ApJ...216..123H, 2014ApJ...782L..10T} and here we focus specifically on the triggering of flares by MHD waves.

\citet{2004A&A...420.1129M}  investigated the behaviour of an aperiodic fast magnetoacoustic pulse about a 2D X-type null point and found that the fast wave refracts into the vicinity of the null point and, ultimately, accumulates at the null point itself. As it approaches the null, the refraction effect causes the pulse amplitude to be amplified and  the length scales (which can be thought of as the distance between the leading and trailing edges of the wave pulse) to rapidly decrease. This leads to an increase of the electric current density associated with the pulse, which manifests as exponential growth  near the null  point. The phenomenon, i.e. fast waves accumulate at null points is entirely general and has been shown to work for double X-type neutral points \citep{2005A&A...435..313M} as well as 3D null points (e.g. \citealt{2012A&A...545A...9T}, {{and see \citealt{2011SSRv..158..205M} for a review}}). Crucially, \citet{2009A&A...493..227M} showed that this accumulation of wave energy at the null  is enough to induce reconnection, i.e. wave-driven reconnection (see \S\ref{sec:OscillatoryReconnection} for full details).

\begin{figure*}
\begin{center}
\includegraphics[scale=0.3]{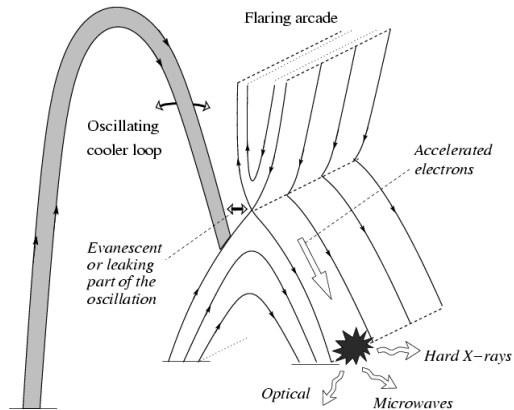}
\caption{The cool (shaded) loop experiences fast magnetoacoustic oscillations (e.g. kink or sausage mode). A segment of the oscillating loop is situated nearby a flaring arcade. An external evanescent or leaking part of the oscillation can reach the null point(s) in the arcade, inducing quasi-periodic modulations of the electric current density. Via plasma micro-instabilities this cause anomalous resistivity which triggers reconnection. This then accelerates particles periodically, which follow the field lines and precipitate in the dense atmosphere, causing quasi-periodic emission in radio, optical and X-ray bands. From \citet{2006A&A...452..343N}.}
\label{Nakariakov2006figure}
\end{center}
\end{figure*}

\citet{2006A&A...452..343N} investigated this phenomenon further by simulating the interaction of a periodic fast magnetoacoustic wave with a magnetic null point. This causes the periodic occurrence of highly steep spikes of the electric current density. The current variations can, in turn, periodically induce current-driven plasma micro-instabilities which are known to cause  anomalous resistivity. This can then {\emph{periodically}} trigger reconnection. The modulation depth of these current variations is a few orders of magnitude greater than the amplitude of the driving wave,  and thus this may be a suitable mechanism for QPPs (see \S\ref{sec:QPPs_section} {{above}}). \citet{2006A&A...452..343N} postulated that this  initial wave driver come from an oscillating coronal loop outside (but close to) the flaring arcade. Thus, an external evanescent or leaking part of the oscillation could reach the null point in the arcade. A sketch of the mechanism can be seen in Figure \ref{Nakariakov2006figure}. Here, the period is determined by the period of the oscillating loop, which corresponds approximately to the ratio of the loop length to the average {{magnetoacoustic}} speed. Moreover, in this scenario the inducing wave may be freely propagating or guided by a plasma non-uniformity, with the periodicity appearing because of its dispersive evolution (see \S\ref{wavetrain} below).

\citet{2006SoPh..238..313C} performed MHD simulations of transition-region explosive events driven by five-minute solar p-mode oscillations. The authors considered {{an anti-parallel magnetic field}} with a stratified atmosphere. Five-minute oscillations are imposed at the photospheric base and this leads to periodically triggered reconnection. Specifically, it was found that density variations in the vicinity of the reconnection site result in a periodic variation in the electron drift speed, which {{switches on/off}} anomalous resistivity, and thus accelerates the process of reconnection {{periodically}}. Thus the reconnection rate is modulated with a period of approximately five minutes. The corresponding UV light curve indicates impulsive bursty behaviour, which each spiky burst lasting for approximately one minute (for the parameters considered). Thus, this  is an example of the periodic triggering of reconnection due to MHD oscillations, specifically longitudinal, slow magnetoacoustic (i.e. p-modes). In addition, this mechanism could readily explain the observed association of QPPs of the microwave emission in solar flares with the slow magnetoacoustic waves leaking from a sunspot in the corona \citep{2009A&A...505..791S}. Thus, both fast and slow magnetoacoustic waves could act as periodic triggers of magnetic reconnection, transferring their periodicities in QPPs.

\subsection{Oscillatory reconnection (Reconnection Reversal)}\label{sec:OscillatoryReconnection} 

\S\ref{periodic triggering} considered periodic flare triggering via MHD oscillations, but alternatively the reconnection itself can be repetitive and even periodic. Traditionally, magnetic reconnection and MHD wave theory have been viewed as separate topics. However, this is a misconception: it is known that steady-state reconnection models generate not only outflows/waves but also require inflows/waves \citep[e.g.][]{1957JGR....62..509P, 1958IAUS....6..123S, 1964NASSP..50..425P}. This point-of-view has been challenged by several authors via the mechanism of {\emph{spontaneous oscillatory reconnection}} which is a time-dependent magnetic reconnection mechanism that naturally produces periodic outputs from aperiodic drivers. From the point of view of oscillation theory, this process is a self-oscillation (see \S\ref{defs} for details). 

% In this way, the unification of magnetic reconnection and MHD waves is a key part of the overarching goal of creating an integrated plasma model of solar flares which will, ultimately, create a coherent vision of reconnection, waves and particle acceleration processes in flares.

The process was first reported  by \citet{1991ApJ...371L..41C} who investigated the relaxation of a 2D X-point magnetic field disturbed from equilibrium. They found that the additional free magnetic energy was released by oscillatory reconnection, which coupled the resistive diffusion at the null point to global advection of the outer field. 

The process is named {\emph{oscillatory}} since inertial overshoot of the plasma carries more flux through the null than is required for equilibrium and the plasma undergoes several oscillations through the null point. The oscillation period scales as $\ln {\eta}$, with $\eta$ being the resistivity. The reconnection rate was found to scale as ${|\ln {\eta}|}^2$, i.e. \lq{fast}\rq{} (as opposed to \lq{slow}\rq{} which depends upon a power of $\eta$). In the theoretical set-up of \citet{1991ApJ...371L..41C}, the free magnetic energy is dissipated across approximately 100 Alfv\'en times and thus, for typical solar parameters, the dissipation time scale is of the order of several minutes to an hour. The authors note that this is sufficiently rapid to account for thermal energy release in the decay phase but may be too slow to explain the impulsive phase flare timescale. This work was expanded upon by \citet{1992ApJ...393..385C}, \citet{1992ApJ...399..159H} and \citet{1993ApJ...405..207C} who suggested that for large-amplitude disturbances the structure of current flattens out into a quasi-1D current sheet. Thus, fast dissipation results in the formation of flux pile-up at the edges of the current layer, and so the bulk of the magnetic energy is released as heat rather than kinetic energy (of bulk mass flows). These works were limited to cylindrical geometries with artificial field line manipulation on the boundary, e.g. imposing a closing-up of the angle of the separatrix field lines, as well as reflective boundaries; thus all the outgoing wave energy was reflected and focused at the null point. Note that the gradients in the  spatially-varying, equilibrium Alfv\'en-speed profile lead to fast magnetoacoustic waves being refracted into the null anyway (see \citealt{2004A&A...420.1129M} for details).

\begin{figure}
\begin{center}
\includegraphics[width=12.0cm]{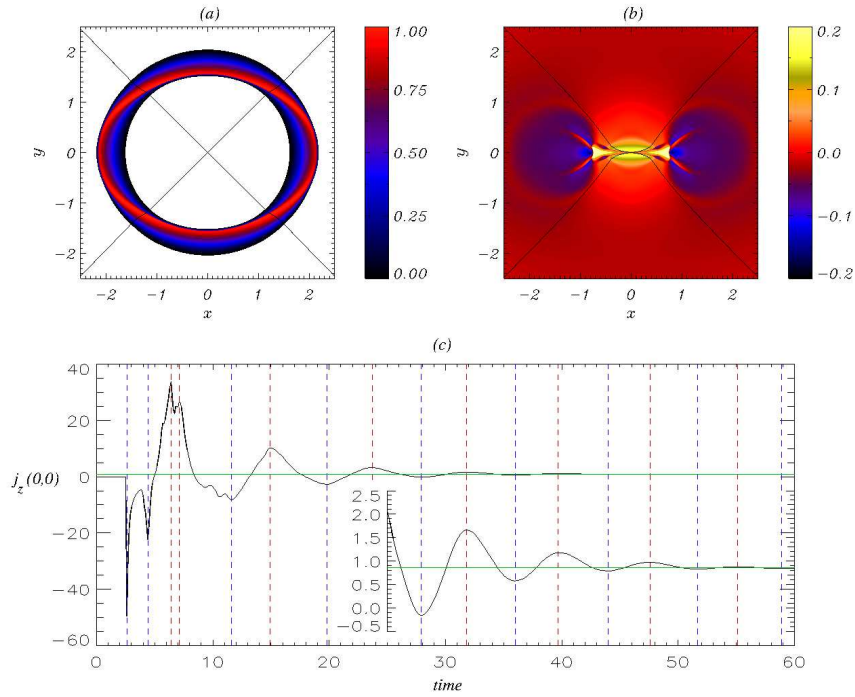}
\caption{Contours of ${\rm{v}}_\perp$ (i.e. velocity across the magnetic field) for a fast wave pulse initially located at a radius $r=5$ and its resultant propagation at $(a)$ time $t=1$ and $(b)$ time $t=2.6$ (measured in Alfv\'en times). Black lines denote the separatrices and null is located at their intersection. Note in $(b)$ the separatrices have been deformed and now form a \lq{cusp-like}\rq{} field structure. Subfigure have their own colour bars since ${\rm{v}}_\perp$  amplitude varies substantially throughout the evolution. $(c)$  Time evolution of $j_z (0,0)$ for $0 \le t \le 60$. Insert shows time evolution of $j_z (0,0)$ for $25 \le t \le 60$ (i.e. same horizontal but different  vertical axis). Dashed lines indicate maxima (red) and minima (blue). Green line shows limiting value of  ${{j_z}(0,0)=0.8615}$. Adapted from \citet{2009A&A...493..227M}.}
\label{figure_OSCILLATORY_RECONNECTION}
\end{center}
\end{figure}

\citet{2009A&A...493..227M} {{were}} the first demonstration of reconnection naturally driven by MHD wave propagation, via the process of oscillatory reconnection. These authors investigated the behaviour of nonlinear fast magnetoacoustic waves near a 2D X-type neutral point and found that the incoming wave deforms the null point into a cusp-like point which in turn collapses to a current sheet.  Specifically, it was found that the incoming (fast) wave propagates across the magnetic field lines and the initial annulus profile contracts as the wave approaches the null. This can be seen in Figure \ref{figure_OSCILLATORY_RECONNECTION}a. The incoming wave was observed to develop discontinuities (for a physical explanation, see Appendix~B of \citealt{2009A&A...493..227M} or, alternatively, \citealt{2011A&A...531A..63G}), and these discontinuities form fast oblique magnetic shock waves, where the shock makes ${\bf{B}}$ refract away from the normal. Interestingly, the shock locally heats the initially plasma $\beta = 0$ plasma, creating plasma $\beta \neq 0$ locally. Subsequently, the shocks overlap and form a shock-cusp, which leads to the development of hot jets and in turn these jets substantially heat the local plasma and significantly deform the local magnetic field (Figure \ref{figure_OSCILLATORY_RECONNECTION}b). When the shock waves reach the null, the initial X-point field has been deformed such that the separatrices now touch one another rather than intersecting at a non-zero angle \citep[called \lq{cusp-like}\rq{ } by][]{1975JPlPh..14..271P}. The osculating field structure continues to collapse; forming a horizontal current sheet. However, the separatrices continue to evolve: the jets at the ends of the (horizontal) current sheet continue to heat the local plasma, which in turn expands. This expansion squashes/shortens the current sheet, forcing the separatrices apart. The (squashed) current sheet thus returns to a \lq{cusp-like}\rq{} null that, due to the continuing expansion of the heated plasma, in turn forms a vertical current sheet. This is the manifestation of the overshoot reported by  \citet{1991ApJ...371L..41C}. The phenomenon then repeats: jets heat the plasma at the ends of this newly-formed (vertical) current sheet, the local plasma expands, the (vertical) current sheet is shortened, the system attempts to return itself to equilibrium, overshoots and forms a (second) horizontal current sheet. The evolution proceeds periodically through a series of horizontal/vertical current sheets. The oscillatory nature  can be clearly seen by looking at the time evolution of the current \citep[in][this was $j_z(0,0)$]{2009A&A...493..227M} as shown in Figure \ref{figure_OSCILLATORY_RECONNECTION}c. We also note that there is nothing unique about the orientation of the first current sheet being horizontal followed by a vertical, this simply results from the particular choice of initial condition, and \citet{2012ApJ...749...30M} use the more general terminology: orientation 1 and orientation 2. \citet{2009A&A...493..227M} also present evidence of reconnection; reporting both a change in field line connectivity as well as changes in the vector potential which directly showed a cyclic increase and decrease in magnetic flux on either side of the separatrices.

\citet{2012A&A...548A..98M} quantified and measured the periodic nature of oscillatory reconnection. They identified two distinct periodic regimes: the (transient) impulsive  phases and a longer-lived stationary phase. In the stationary phase, for driving amplitudes $6.3-126.2$ km/s, they measured (stationary-phase) periods in the range $56.3-78.9$ s. In particular, a driving amplitude of $25.2$ km/s corresponds to a stationary period of $69.0$ s. \citet{2012A&A...548A..98M} highlighted that the system acts akin to a damped harmonic oscillator (in the stationary phase). Thus, the greater the initial amplitude, the longer and stronger the current sheets at each stage, and thus the greater restoring force, leading to shorter periods (compared to smaller initial amplitude, shorter resultant current sheets, weaker restoring force and thus longer periods).

The physics behind oscillatory reconnection has been investigated by \citet{2009A&A...493..227M}, \citet{2009A&A...494..329M} and  \citet{2012A&A...544A..24T}. The restoring force of oscillatory reconnection has been shown to be a dynamic competition  between the thermal-pressure gradients and the Lorentz force (i.e. a local imbalance of forces)  with each in turn restoring an overshoot of the equilibrium brought on by the other  (see  Sect. 3.3 of \citealt{2009A&A...493..227M}; Sect. 3.2 of  \citealt{2009A&A...494..329M};  Fig. 7 of \citealt{2012A&A...544A..24T}). In other words, the  reconnection occurs in distinct bursts: the inflow/outflow magnetic fields of one reconnection burst  become the outflow/inflow fields in the following burst. With the Lorentz force, it is magnetic pressure that dominates  \citep{2012A&A...544A..24T}  whereas magnetic tension only aids the compression of field lines as the current sheet forms. Note that in consecutive bursts of reconnection, the contrast between the thermal-pressure gradients and magnetic pressure  decreases. Thus, each successive overshoot is smaller than the last and so the system is ultimately able to relax back to equilibrium. 

\subsubsection{Periodic signals associated with magnetic flux emergence}

An important example of oscillatory reconnection was found  in the  work  of \citet{2009A&A...494..329M}  who utilised a stratified atmosphere permeated by a unipolar magnetic field (to represent a coronal hole) and investigated the emergence of a buoyant flux tube. Flux emerging into a pre-existing field had been studied in  detail before, but \citet{2009A&A...494..329M}  were the first to investigate the long-term evolution of such a system, i.e. previous simulations ended once reconnection was first initiated. Murray et al. found that a series of \lq\lq reconnection reversals" take place as the system searches for equilibrium, i.e. a cycle of inflow/outflow bursts followed by outflow/inflow bursts. Thus, the system demonstrates oscillatory reconnection in a self-consistent manner. 

This seminal work was generalised by  \citet{2012ApJ...749...30M} who detailed the oscillatory outputs and outflows of the system. They found that the physical mechanism of oscillatory reconnection naturally generates quasi-periodic vertical outflows with a transverse/swaying aspect. The vertical outflows consist of both a periodic aspect and  a positively-directed flow of $20-60$~km/s. Parametric studies show that varying the magnetic strength of the initial-submerged, buoyant flux tube  ${\bf{B}}_{\rm{buoyant}}$ yield a  range of associated periodicities of $105-212.5$~s for $2.6 \times 10^3$~G $\le {\bf{B}}_{\rm{buoyant}} \le$ $3.9 \times 10^3$~G, where the stronger the initial flux tube strength, the longer the period of oscillation. Note that if the flux tube strength is too low  \citep[for][this was ${\bf{B}}_{\rm{buoyant}}  < 2.6 \times 10^3$ G]{2012ApJ...749...30M} the tube cannot fully emerge into the atmosphere since the buoyancy instability criterion is not satisfied (failed emergence). If the initial-submerged flux was too high (${\bf{B}}_{\rm{buoyant}} > 3.9 \times 10^3$~G) then plasmoids are ejected from the ends of the current sheet. These ejected plasmoids change the properties of the X-point, e.g. taking magnetic flux with them. Thus, even though there is still oscillatory behaviour, this represents a fundamentally different regime than that of burst of reconnection without plasmoids. Thus, there are natural limitations placed on the periods generated by oscillatory reconnection in such a system. As before, the  mechanism  naturally generates periodic outputs even though no periodic driver is imposed on the system. Note that the transverse behaviour seen in the periodic jets originating from the reconnection region of the inverted Y-shaped structure is specifically due to the oscillatory reconnection mechanism, and would be absent for a single, steady-state reconnection jet. 

\begin{figure}
  \centering
    \includegraphics[width=120mm]{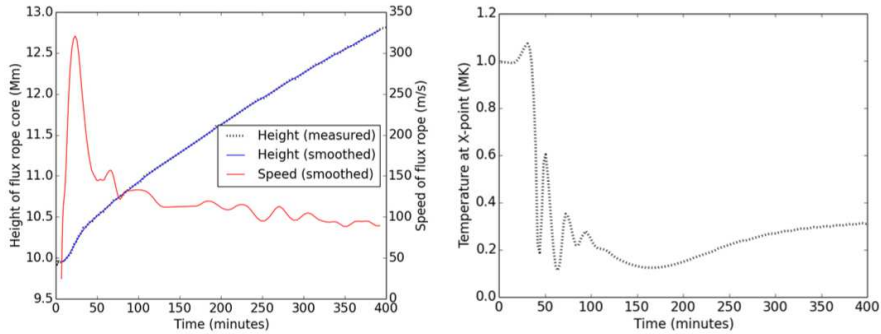}
 \caption{Variations of the magnetic rope speed and temperature at the X-point below the flux rope, caused by the interaction of an emerging magnetic flux with a pre-existing magnetic configuration.  From \citet{2014A&A...569A..94L}.}
    \label{lukin14}
\end{figure}

Thus, oscillations associated with magnetic flux emergence (as well as the continuous emergence of the magnetic flux) show promise as a physical mechanism for QPPs, for example  \citet{2012ApJ...749...30M} could not generate periodicities shorter than $105$ seconds since this was restricted by the buoyancy instability criterion (i.e. failed emergence): lower periods could have been generated by changing the equilibrium parameters, such as modifying the strength of the pre-existing coronal field in the model. Longer periods are also possible for different equilibrium set-ups, e.g. \citet{2014A&A...569A..94L} saw 30-min oscillations during the interaction of an emerging magnetic flux with a pre-existing coronal magnetic configuration, see Fig.~\ref{lukin14} (Figure 9 from \citealt{2014A&A...569A..94L}). 

\subsubsection{Periodicities generated}

\citet{2012A&A...548A..98M} also ask what  determines the (stationary) period (post-transients) and what determines the exponentially-decaying timescale.  They recall the work of \citet{1991ApJ...371L..41C} who derived an analytical prediction for two timescales:
\begin{equation*}
t_{\rm{oscillation}} \approx 2 \ln {S}\;,\;\; t_{\rm{decay}}\approx {t_{\rm{oscillation}}}^2 / {2 \pi^2}
\end{equation*}
where $S$ is the Lundquist number and we identify $t_{\rm{oscillation}}$ as our stationary period and $t_{\rm{decay}}$ as our decay time. From \citet{2012A&A...548A..98M}, for a driving amplitude of $25.2~$km\,s$^{-1}$, this gives  $t_{\rm{oscillation}} = 109.6~$s and $t_{\rm{decay}}=76.7~$s, compared to the  measured stationary period of $69.0~$s and decay time of $66.7~$s. These estimates are in fair agreement given the simplicity of the \citet{1991ApJ...371L..41C} model {{and hence we could}} conclude that the periodicity is determined by the Lundquist number, $S$, or equivalently the magnetic Reynolds number, $R_m$, since everything in \citet{2012A&A...548A..98M} is non-dimensionalised with respect to the Alfv\'en speed.

{{
However, the model of \citet{1991ApJ...371L..41C}  is a closed system, whereas  \citet{2012A&A...548A..98M} is an open system. Thus, the period of  \citet{1991ApJ...371L..41C} is better interpreted as the signal travel time from their outer boundary to the diffusion region (see, e.g. $\S7.1$ of \citealt{2000mare.book.....P} for further discussion). \citet{1991ApJ...371L..41C} also neglect both nonlinear and thermal-pressure effects and so,  in that sense, the similarity in periods between \citet{1991ApJ...371L..41C} and \citet{2012A&A...548A..98M} could be simply coincidental. Thus, {\it{what dictates the period of oscillatory reconnection}} remains an open question.
}}

%where $S$ is the Lundquist number and we identify $t_{\rm{oscillation}}$ as our stationary period and $t_{\rm{decay}}$ as our decay time. From \citet{2012A&A...548A..98M}, for a driving amplitude of $25.2~$km\,s$^{-1}$, this gives  $t_{\rm{oscillation}} = 109.6~$s and $t_{\rm{decay}}=76.7~$s, compared to the  measured stationary period of $69.0~$s and decay time of $66.7~$s. These estimates are in fair agreement given the simplicity of the \citet{1991ApJ...371L..41C} model which neglected both nonlinear and thermal-pressure effects. Thus, we {{could}} conclude that the periodicity is determined by the Lundquist number, $S$, or equivalently the magnetic Reynolds number, $R_m$, since everything in \citet{2012A&A...548A..98M} is non-dimensionalised with respect to the Alfv\'en speed. 

The periodicities generated by the oscillatory reconnection mechanism are promising: around a lone null point, periodicities of $56.3-78.9$ s have been found, and via flux emergence scenarios, periodicities of $105-212.5$ s have arisen in a self-consistent manner. Flares  unlock the stored non-potential magnetic energy in magnetic fields and, by releasing energy, a stressed magnetic system can return to a lower energy state. As noted by \citet{2009A&A...494..329M}  flares, therefore, are perfect events in which to search for signs of oscillatory reconnection.

Crucially for the mechanism  these oscillations are {\emph{generated}} with an exponentially-decaying signature for both the flux emergence scenario   \citep{2012ApJ...749...30M} and single null \citep{2012A&A...548A..98M}. QPPs with these properties have been detected in soft X-ray light curves of both solar and stellar flares (e.g. \citealt{2016ApJ...830..110C}{\footnote{{{Note that \citet{2016ApJ...830..110C} concluded the mechanism of $\S\ref{sec:coronal_seismology}$, rather than that of $\S\ref{sec:OscillatoryReconnection}$,  was more applicable to their observations.}}}}). {{What is important to note is that}} the oscillations caused by this mechanism would be decaying not due to a particular dissipative mechanism, but {\emph{due to the generation mechanism itself}}. Physically, this can be thought of as injecting a finite amount of energy into the oscillatory reconnection mechanism and so, intuitively, the resultant periodic behaviour must be finite in duration. Clearly this is a dynamic reconnection phenomena as opposed to the classical steady-state, time-independent reconnection models. This means that the oscillatory reconnection mechanism will struggle to explain decayless oscillations.

Only  specific examples of oscillatory reconnection  have been investigated so far but, given that the underlying physical mechanism in the  dynamic competition between gas and magnetic pressure searching for equilibrium,  the mechanism looks to be a robust, general phenomenon that may be observed in other systems that demonstrate finite-duration reconnection. The mechanism could occur at all scales. {{Recently, \citet{2017ApJ...844....2T} demonstrated how the oscillatory reconnection mechanism works about a three-dimensional null point, and now parametric }} studies are needed to investigate the full range of periodicities possible, as well as an investigation into how plasmoid generation modifies the system.  Further studies should focus on  heat conduction which is expected to reduce the temperature of the outflow jets. However, to ensure force balance in the current sheet, the density of outflows may actually be increased by heat conduction, which may make the outflow jet more observable. Another interesting question is whether this mechanism can produce QPPs in flaring light curves, if in the flare site there are several or a number of plasmoids and hence, elementary null points, as has been suggested in the fractal reconnection model \citep{2016ASSL..427..373S}.

\subsection{Thermal over-stability}\label{theros}
      
\begin{figure}[htp]
  \centering
  \begin{tabular}{cc}
    \includegraphics[width=5.0in]{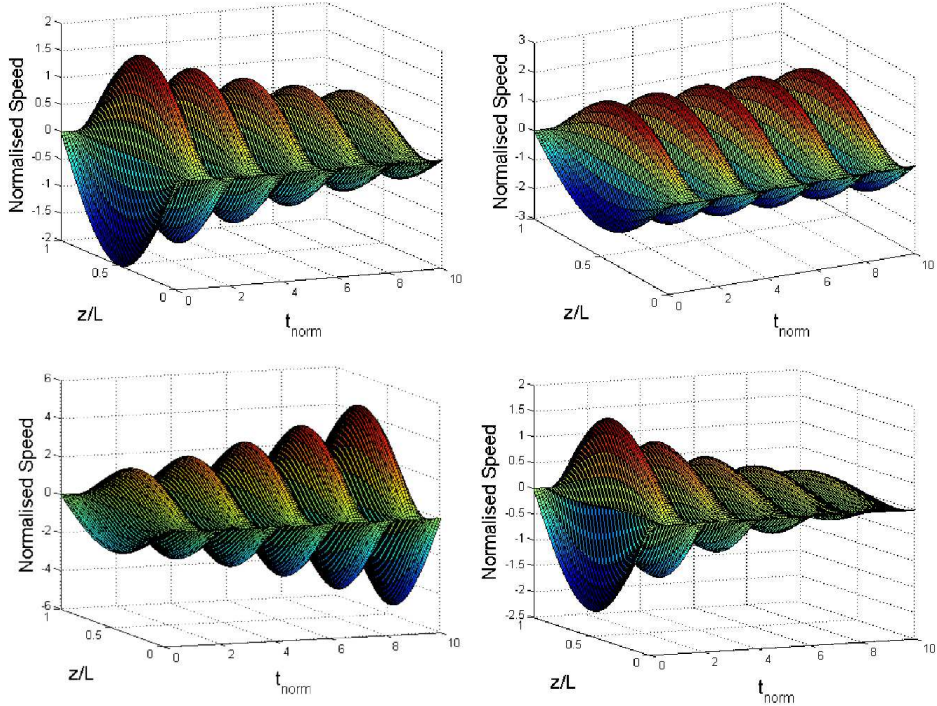}
  \end{tabular}
 \caption{Different regimes of the evolution of the fundamental acoustic oscillation of a coronal loop, determined by the misbalance of radiative cooling and plasma heating. The plasma speed is normalised at double the initial amplitude. The time is normalised at half the oscillation period. The spatial coordinate is normalised at the loop length $L$. 
Top raw, left: a decaying linear oscillation in the absence of radiative cooling and heating; right: undamped oscillation occurring when radiative and dissipative losses are compensated by heating.
Bottom raw, left:  a growing oscillation; 
right: an over-damped oscillation. Adapted from \citet{2016ApJ...824....8K}.}
    \label{kumar2016}
\end{figure}

In the solar coronal plasma there is a continuous competition between the radiative losses and the energy supply, that constitutes the coronal heating problem \citep[see, e.g.][for a recent review]{2012RSPTA.370.3217P}. The misbalance of the radiative losses and heating can lead to the appearance of oscillatory regime of thermal instability, and variations of thermodynamical properties of the plasma and induced flows. In particular, the dispersion relation describing acoustic oscillations along the field is: 
\begin{equation}
\omega^2-C_\mathrm{s}^2 k^2-  \frac{i(\gamma - 1)}{\rho_0} \Big(\frac{\rho_0  \tilde{a_\rho} k^2}{\omega} + \tilde{a_p} \rho_0 \omega  
- \frac{\overline{\kappa}  m  \omega  k^2}{k_B}+\frac{\overline{\kappa} T_0 k^4}{\omega}\Big) +i\overline{\nu}\omega k^2 = 0 ,
\label{ti_disprel}
\end{equation}
where $\omega$ is the cyclic frequency, $k$ is the wave number, $C_\mathrm{s}$ is the sound speed, $\overline{\kappa}$ and $\overline{\nu}$ are the parallel thermal conductivity and bulk viscosity, respectively, and
$\tilde{a_\rho}={\partial Q}/{\partial \rho}$ and $\tilde{a_p}={\partial Q}/{\partial p}$ are the derivatives of the combined plasma heating/cooling function $Q(p,\rho)$ at the thermal equilibrium with the pressure $p_0$ and density $\rho_0$, and other notations are standard \citep[see][for details]{2016ApJ...824....8K}. The heating mechanism is not specified, and is assumed to be stationary. The radiation is assumed to be optically thin. In the case of weakly non-adiabatic effects, one can readily separate the real and imaginary parts of dispersion relation (\ref{ti_disprel}), obtaining:
\begin{eqnarray}
 \cal{R}(\omega)&\approx&C_\mathrm{s} k, \label{omegar} \\ 
 \cal{I}(\omega) &\approx& \frac{(\gamma - 1)}{2}\Big( \tilde{a_\rho}/C_\mathrm{s}^2+\tilde{a_p}\Big)   
 -\Big[\frac{(\gamma - 1)^2 m\overline{\kappa}}{2\gamma \rho_0 k_B}+ \frac{\overline{\nu}}{2}\Big]k^2, \label{omegai}
\end{eqnarray}
respectively. The sign and specific value of $\tilde{a_\rho}/C_\mathrm{s}^2+\tilde{a_p}$ is determined by the dependence of the radiative and heating function on thermodynamical variables. If the value is positive, the misbalance of the radiative losses and plasma heating counteracts the dissipation because of thermal conduction and viscosity. Moreover, if this value is sufficiently large, the acoustic oscillation becomes undamped and even growing (see Figure~\ref{kumar2016}). In the case of the negative value, this effect enhances the oscillation damping. In the over-stable regime, the oscillation amplitude experiences the saturation because of nonlinear effects. The oscillation frequency is determined by the wavelength, e.g. the distance between the footpoints along the magnetic field line \citep[e.g.][]{2004A&A...422..351T}. This phenomenon is acoustic over-stability that can occur in flaring regions. As the second term on the right hand side of Eq.~(\ref{omegai}) depends on $k^2$, the acoustic over-stability is most pronounced for long wavelength perturbations, for example, fundamental modes of long loops. Typical periods of the quasi-periodic pulsations of thermal emission, generated by acoustic self-oscillations, are determined by the length of the oscillating loop and the plasma temperature. For typical flaring loops the periods of these oscillations range from a few to several minutes, and may be longer in the case of these oscillations in long, cold pre-flare loops or filaments. 

Examples of the decayless and growing oscillations detected in the Doppler shifts of hot coronal emission lines (possibly, incorrectly best-fitted by a decaying harmonic function) could be seen in Figure~3 of \citet{2008ApJ...681L..41M}. The recently detected very long period pulsations of the plasma temperature before the onset of flares \citep[8--30 min \lq\lq preflare-VLPs\rq\rq,][]{2016ApJ...833..206T} may perhaps be linked to this effect too. In addition, this effect may be responsible for the high-quality oscillatory patterns of the thermal X-ray emission, with the intermittent variation of the amplitude, detected in the time derivative of the GOES light curves of X-class flares by \citet{2015SoPh..290.3625S} {{(see also \citealt{2016ApJ...827L..30H} and \citealt{2017ApJ...836...84D}).}}

An interesting research avenue is the investigation of the effect on the misbalance between the radiative losses and (quasi-)steady heating on another highly compressive mode, the sausage oscillation. Damping of this mode is known to be connected with the finite transport coefficients, i.e. the ion viscosity and electron  thermal conductivity, and also by leakage of the fast magnetoacoustic oscillations across the field, in the external medium \citep[e.g.][]{2014Ge&Ae..54..969S, 2012ApJ...761..134N}. On the other hand, observations show the presence of high-quality compressive QPP with the periods typical for the sausage mode. For example, 25-s intensity and Doppler shift oscillations were recently detected {{in}} the thermal emission by \citet{2016ApJ...823L..16T}, and interpreted as the sausage mode. In these oscillations, the dissipative, radiative and leaky losses should be compensated by some energy supply that could be the thermal over-stability.

Longer period variations of the thermal emission intensity could be associated with entirely thermodynamical processes, for example the evaporation-condensation cycles
\citep[e.g.][and references therein]{2017ApJ...835..272F}. These cycles include recurrent plasma condensations and temperature variations, followed by high-speed plasma flows \citep{2004A&A...424..289M}, even in the case of a time-independent heating function. In this regime the QPP patterns are usually highly anharmonic, and resemble relaxation oscillations. It is found that this effect gives a wide range of periods, while we are not aware of any systematic studies of the dependence of the oscillation period upon the plasma parameters. Similar quasi-oscillatory variations are observed in laboratory plasma devices, in particular the phenomenon of the multifaceted asymmetric radiation from the edge \citep[\lq\lq MARFE\rq\rq,][]{deploey}.

\subsection{MHD flow over-stability}\label{waveflow}  
 
In the self-oscillation scenario, the energy supply can also be associated with steady flows of the plasma. \citet{2006ApJ...644L.149O} considered the dynamical reconnection in a current sheet with a steady plasma flow localised near its plane. The profiles of all equilibrium quantities were taken to be smoothly non-uniform in the transverse direction. The profile of the flow had the transverse spatial scale about one order of magnitude smaller than the transverse non-uniformity of the magnetic field. The plasma density and temperature was initially constant. In the vicinity of the current sheet the plasma $\beta$ was taken to be high, of about 4. Such a plasma configuration could appear because of, for example, the interaction between an emerging flaring loop and the overlying magnetic field. In this scenario, the plasma flow is caused by the chromospheric evaporation, which can be taken as steady if its time scale is much longer than the period of QPPs. 

\begin{figure}
  \centering
    \includegraphics[width=3.0in]{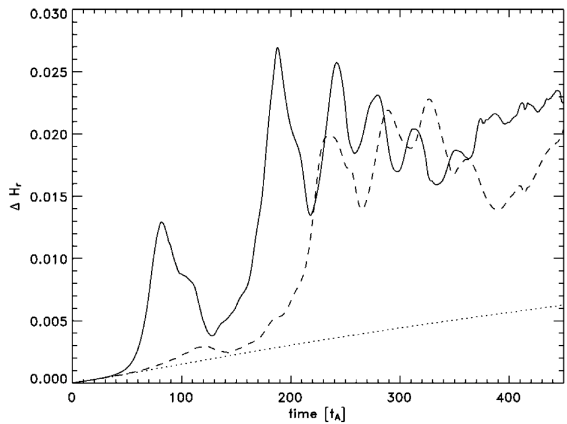}
 \caption{Variation of the integrated Ohmic heating rate, $\Delta H_r$, in a reconnecting current sheet with a non-uniform steady flow of the plasma. The dotted curve corresponds to the case without the flow, the dashed curve shows the case when the flow speed is equal to the local Alfv\'en speed, and the solid line to the case when the flow speed is 1.5 of the local Alfv\'en speed. The time unit is the transverse Alfv\'en time that is the ratio of the current sheet half-width and the Alfv\'en speed. From \citet{2006ApJ...644L.149O}.}
    \label{ofman2006}
\end{figure}

For a sufficiently high speed of the plasma flow, e.g. about the Alfv\'en speed, the plasma configuration was found to be unstable to the coupled Kelvin--Helmholtz and tearing instabilities, giving rise {{to}}  the over-stable, i.e. oscillating, modes. During the evolution, the integrated Ohmic heating rate was found to vary quasi-periodically, see Figure~\ref{ofman2006}. The oscillation period is about 50 Alfv\'enic transit times across the current sheet. In the numerical simulations of \citet{2006ApJ...644L.149O}, for a macroscopic current sheet of the half-width about 1,500 km, and the Alfv\'en speed of 500 km/s, the oscillation period is about 150~s.

This mechanism can naturally produce QPPs of the thermal emission, by the variation of the plasma heating rate. In addition, as the over-stability leads to the development of magnetic islands (plasmoids), there appear strong oscillating electric field that can readily exceed the Dreicer field. Hence, the over-stability is accompanied by the periodic acceleration of non-thermal electrons and associated QPP of non-thermal emission. 

A parametric study of this mechanism, in particular the investigation of the effect of the specific values of the transport coefficients, the steepness of the transverse profiles of the flow, the electric current and plasma densities, temperature, magnetic field, and the anomalous resistivity, on the oscillation period, would be an interesting future task. Also, the over-stability could be associated not with the Kelvin--Helmholtz instability, but with one of the negative energy instabilities that have a much lower shear flow threshold, e.g. \citet{1997SoPh..176..285J}.

\subsection{Waves and plasmoids in a current sheet}\label{wavetrain}

Neutral current sheets are common structures both in the solar corona and in magnetosphere of the Earth. In the solar atmosphere, we can find these structures, for example, above the helmet structures, in coronal streamers, at the boundary between closed and open fields, or between coronal loop systems which have opposite magnetic polarity. A macroscopic current sheet is the key ingredient of the standard model of a solar flare. Neutral current sheets can be formed in one of three possible ways: the interaction of topologically different regions (this may give rise to a solar flare), the loss of equilibrium of a force-free field, and X-point collapse {{(see \citealt{2014masu.book.....P})}}. Owing to their enhanced density, neutral current sheets are the structures that can guide the MHD waves. A Harris-type current sheet structure can support several kinds of guided magnetoacoustic waves, in particular, both kink and sausage modes. In the sausage mode, the current sheet pulsates like a blood vessel, with the central axis remaining undisturbed. In the kink mode the central axis moves back and forth during the wave motion. In a continuously non-uniform current sheet, three types of mode can exist: body, surface and hybrid, depending on the transverse structure of the perturbation \citep{1997A&A...327..377S}. Hybrid modes contain elements of both body and surface waves, see e.g. \citet{1994JPlPh..51..221C}, \citet{1997A&A...327..377S} and references therein. The nature of the mode determines its dispersion, i.e. the dependence of the phase speed on wavenumber.  MHD waves and oscillations of current sheets can be excited by various processes where one of the most probable, providing either single or multiple sources of disturbances, is the impulsive energy release in a flare. In turn, these oscillations, e.g. the quasi-periodic signals appearing because of the dispersion, can be the sources of QPPs. 

The analysis of group speeds of the guided modes shows that the long-wavelength spectral components propagate faster than the medium- and short-wavelength ones. This suggested that an impulsively-generated fast wave train has a characteristic wave signature with three distinct phases: the periodic phase, followed by quasi-periodic phase and then a decay phase \citep{1983Natur.305..688R,1984ApJ...279..857R}. \citet{2004MNRAS.349..705N} simulated the formation of a quasi-periodic wave train in fast magnetoacoustic waveguides with the transverse plasma density profiles of different steepness. It was established that the dispersive evolution of fast wave trains leads to the appearance of characteristic ``crazy" tadpole wavelet signatures, which was also confirmed by the observations. The key element of this mechanism is the broadband excitation, in other words, by a pulse that could occur because of a flaring energy release. 

\citet{2013A&A...560A..97P}  simulated numerically the dispersive evolution of fast waves in an expanding magnetic field generated by an impulsive, spatially localised energy release with a field-aligned density structures. The numerical results were found to be consistent with the observations, see \citet{2013A&A...554A.144Y}. \citet{2013MNRAS.434.2347J} numerically studied magnetoacoustic-gravity waves in an open magnetic structure. They found that a pulse of the horizontal velocity both below and above the transition region could trigger oscillations with the periods in the range of three minutes, which correspond with those observed above the sunspots e.g. in UV/EUV emission by the Solar Dynamics Observatory (SDO)/Atmospheric Imaging Assembly (AIA) and in radio emission by the Nobeyama Radioheliograph (NoRH). The propagation of fast magnetoacoustic waves along coronal magnetic funnels has been studied numerically in \citet{2015ApJ...800..111Y}. The waves are excited impulsively by plasmoids formed in the X-point in the coronal magnetic funnel structure followed by the collision between them and the magnetic field in the outflow region. \citet{2014A&A...569A..12N} found good agreement of the numerical simulations of rapidly propagating fast wave trains with the observations of quasi-periodic rapidly-propagating waves of the EUV intensity observed with SDO/AIA. It was found out that an impulsive energy release could generate a quasi-periodic propagating fast wave train with a high signal quality from a single impulsive source. All the above mentioned studies modelled the fast wave propagation in a plasma slab. However, the results are not sensitive to the direction of the magnetic field as long as it is parallel to the slab's boundaries. Hence, these results could be applied to the case of a neutral current sheet, provided it remains stable on the time scale of the wave evolution. 

The widely used current sheet model satisfying the MHD equilibrium, $\nabla p = \mathbf{j} \times \mathbf{B}$, is the so-called Harris model given by the magnetic configuration: 
\begin{equation}\label{eq1}
\mathbf{B} = B_{\mathrm{0}} \tanh \left(\frac{y}{w_\mathrm{cs}}\right) \hat{\mathbf{e}}_x,
%\mathbf{B} = B_{\mathrm{0}} \tanh \left[\frac{(y - H/2)}{w_\mathrm{cs}}\right] \hat{\mathbf{e}}_x,
\end{equation}
where $B_{\mathrm{0}}$ is external magnetic field and $w_\mathrm{cs}$ is the semi-width of the current sheet. This formula was first derived by \citet{Harris1962} in terms of the kinetic Vlasov theory.

It is well known that magnetoacoustic waves can be triggered easily during reconnection of magnetic field lines. In \citet{2000A&A...360..715K} the authors present a 2D MHD numerical model of pulsating decimetric continuum radio bursts, caused by quasi-periodic particle acceleration, resulting from the dynamic phase of magnetic reconnection in a large-scale current sheet. By means of this model, where the formation of plasmoids, their coalescence and repeated formation of next plasmoids, they explain the presence of quasi-periodic pulses with the characteristic periods ranging in $0.5-10~\mathrm{s}$.

Radio spikes, defined as a group of very short and narrowband bursts, are observed during solar flares and are believed to be generated during the reconnection process \citep{2001A&A...379.1045B}. \citet{2011A&A...529A..96K, 2011ApJ...737...24B} demonstrated that narrowband dm spikes could be associated with fast magnetoacoustic waves, numerically modelling the waves excited by turbulent reconnection outflows in a neutral Harris current sheet. The dispersively evolved waves were found to have the same wavelet spectral signatures as detected in the radio observations. It was concluded that narrowband dm spikes are generated by driven coalescence and fragmentation processes in turbulent reconnection outflows. The propagating magnetoacoustic waves (indicated by tadpole wavelet spectral signatures) modulate these coalescence processes via a modulation of current densities in interaction regions between colliding plasmoids. These waves modulate an acceleration of electrons and generation of plasma and electromagnetic waves that produce the spikes. The narrowband dm spikes can thus be considered as a radio signature of the fragmented reconnection in solar flares.

\begin{figure*}
  \hspace*{-1.5cm}
\includegraphics[scale=0.29]{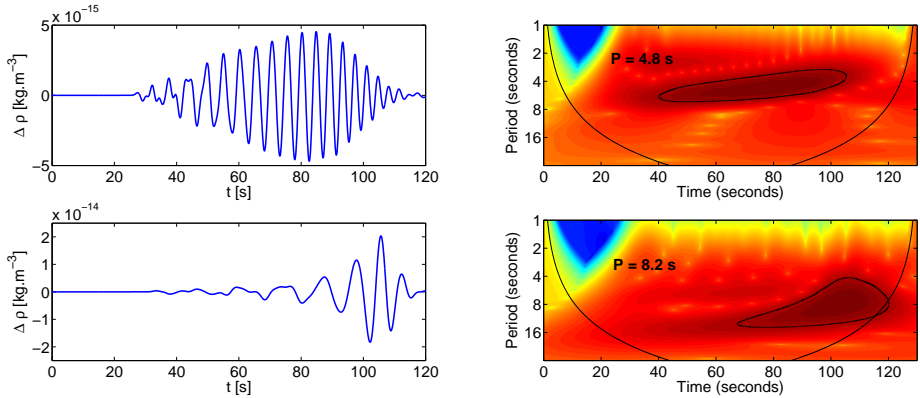}
\caption{Comparison of wave signals (left column) and corresponding wavelet tadpole shapes (right column) for two different widths of the Harris current sheet; $w_{\mathrm{cs}} = 0.50~\mathrm{Mm}$ (first row) and $w_{\mathrm{cs}} = 1.50~\mathrm{Mm}$ (second row).  From \citet{2012A&A...537A..46J}.}
\label{fig2}      
\end{figure*}

\citet{2012A&A...537A..46J} performed a more extended and detailed study of fast sausage waves in a current sheet. The specific interest has been placed on the parameters of the current sheets, such as the width, plasma $\beta$, and the distance between the wave initiation and detection sites, that influence the detected signal and its corresponding wavelet spectrum, see Fig. \ref{fig2}. The wave period, similarly as in the case of simple mass density slab, can be expressed as:
\begin{equation}\label{eq2}
P \approx \frac{w_\mathrm{cs}}{v_\mathrm{Ae}},
\end{equation}
where $v_\mathrm{Ae}$ is the external Alfv\'{e}n speed.

%The data were recorded in the position $L_\mathrm{D} = \{L/2;H/2\}$.

Assuming that fast magnetoacoustic waves guided by current sheets modulate the radio fluxes (or even UV fluxes) at various locations,  \citet{2012A&A...537A..46J} proposed that this knowledge can be helpful for estimating physical parameters of flare current sheets --- another example of MHD seismology. From the point-of-view of the diagnostics of either flare current sheets or flare loops, the most important measurements and findings were: a) the periods of the fast waves, which give information about the half-width of the Harris current sheet, and b) that the wavelet tadpoles become longer and their heads are detected later in time when increasing the distance between the detection and perturbation points. Thus, it is possible to estimate the distance between the radio source at which the modulated signal is detected, and the region where the modulating magnetoacoustic wave is initiated. For example, the magnetoacoustic wave can propagate along the current sheet upwards in the solar atmosphere, and modulate the radio emission (produced by the plasma emission mechanism) at lower radio frequencies. The wavelet spectra of the signals at these frequencies would then show how the wavelet tadpoles have shifted in time, corresponding to the propagating magnetoacoustic wave train. Each tadpole corresponds to a specific plasma frequency, i.e. to the specific plasma density. Using models of the density stratification it is possible to determine the height. In particular, the recent detection of a quasi-periodic sequence of short radio \lq{sparks}\rq{} (finite-bandwidth, short-duration isolated radio spikes) revealed that their repetition rate, 100~s, coincides with the periodicity in a quasi-periodic rapidly-propagating train of the EUV emission, detected in the low corona \citep{2016A&A...594A..96G}.  

There also exist several important differences between the propagation of fast magnetoacoustic waves in a vertical flare current sheet in a gravitationally-stratified solar atmosphere and a gravity-free case. {{The authors \citep{2002A&A...383..685G,2012A&A...546A..49J} implemented in their 2D numerical simulations for the altitude-variant current sheet in the gravitational-stratified solar atmosphere an additional horizontal component of the magnetic field, contrary to the gravity-free case and altitude-invariant current sheet:}}
\begin{equation}\label{eq3}
B_x(x,y) = B_{\mathrm{0}} \frac{w_\mathrm{cs}}{H_0} \ln \left[\cosh \left(\frac{x}{w_\mathrm{cs}}\right)\right] \exp \left(-\frac{y}{H_0}\right),
\end{equation}

\begin{equation}\label{eq4}
B_y(x,y) = B_{\mathrm{0}} \tanh \left(\frac{x}{w_\mathrm{cs}}\right)\exp \left(-\frac{y}{H_0}\right),
\end{equation}
where the coefficient $H_0$ denotes the magnetic scale height.

\begin{figure*}
\hspace*{-1.5cm}
  \includegraphics[scale=0.29]{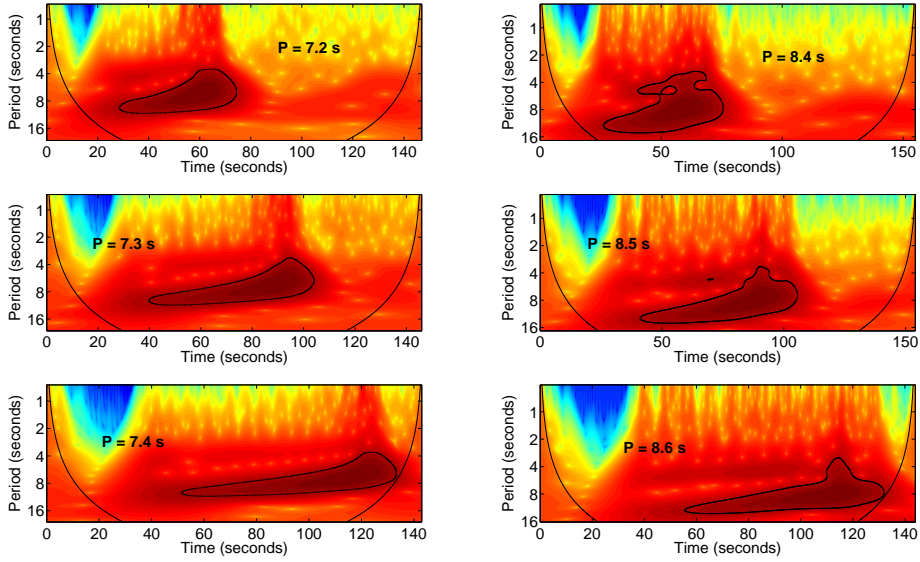}
\caption{Temporal evolution of wavelet tadpoles for the three detection points: $L_\mathrm{D} = 50~\mathrm{Mm}$, $60~\mathrm{Mm}$, and $70~\mathrm{Mm}$ (upper, middle, and lower panel, respectively) for the gravity-free (left panels) and gravitationally stratified (right panels) solar atmosphere. The semi-width of the current sheet is $w_\mathrm{cs} = 1.0~\mathrm{Mm}$. From \citet{2012A&A...546A..49J}.}
\label{fig4}     
\end{figure*}

As a consequence of this modification, waveguiding properties of the current sheet can change significantly {{\citep{2012A&A...546A..49J}}}. At very low altitudes of the vertical current sheets the parameters are the same in both cases. However, in the stratified case the width of the current sheet grows with height \citep{2012A&A...546A..49J}. {{By this fact the authors in their 2D numerical simulations explained (according to Eq. \ref{eq2}) the longer wave periods of propagating fast magnetoacoustic waves in the gravitationally-stratified solar atmosphere compared to the gravity-free case, see Fig. \ref{fig4}.}}

Variations of the wave signal and their wavelet tadpoles are more irregular in the case with gravity {{(altitude-variant current sheet)}} than in the gravity-free case {{{(altitude-invariant current sheet)}}}, which result from the variation with height of the dispersive properties and group velocities of the propagating magnetoacoustic waves in the gravitational case. As the gravitationally-stratified atmosphere is more realistic than  gravity-free,  it allows one to make a direct comparison with observational data. The most frequently measurable parameters of these waves in solar events are the wave periods and their temporal changes (i.e. the period modulation). Combining these data with the possible determination of the wave types and their wavelengths (from spatially-resolving measurements) together with independent estimates of the Alfv\'en speed at these locations (e.g. by the magnetic field extrapolation or UV and optical spectroscopy methods), it could be possible to directly compare these results with the observational findings.

%Fiber bursts are fine structures of broadband type IV radio bursts, manifested by a certain frequency drift. In the wavelet spectra of the fiber bursts computed at different radio frequencies, wavelet tadpole features were found, whose head maxima have the same frequency drift as the drift of fiber bursts, see Fig. \ref{fig5}.

\begin{figure*}
\begin{center}
  \includegraphics[scale=0.5]{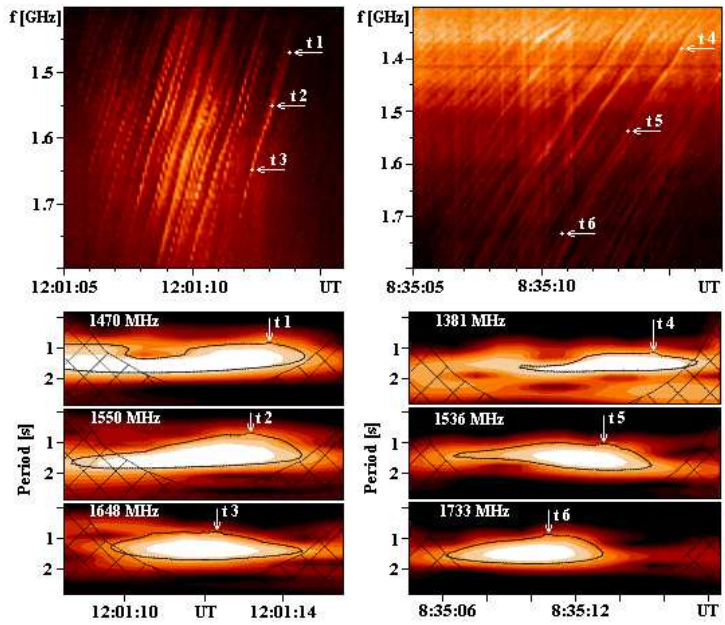}
\caption{Upper panels: examples of the fiber bursts at 12:01:05--12:01:16 UT (November 23, 1998, left panel) and 8:35:05--8:35:17 UT (November 18, 2003, right panel). Bottom part: corresponding wavelet power spectra showing the tadpoles with the period $P = 1.4~\mathrm{s}$. In both spectra, at selected frequencies the times of the tadpoles head maxima were determined ($t1-t6$) and also shown in the upper dynamic radio spectra (upper panels). The tadpole head maxima drift as the fiber bursts. From \citet{2013A&A...550A...1K}.}
\label{fig5}       
\end{center}
\end{figure*}

Fiber bursts are fine structures of broadband type IV radio bursts, manifested by a certain frequency drift. In the wavelet spectra of the fiber bursts computed at different radio frequencies, wavelet tadpole features were found, whose head maxima have the same frequency drift as the drift of fiber bursts, see Fig. \ref{fig5}. It indicates that the drift of these fiber bursts can be explained by the propagating fast sausage wave train, which modulates the radio emission produced by non-thermal electrons trapped in a flare loop or current sheet. \citet{2013A&A...550A...1K} presented a model for fiber bursts in the dm band, based on assuming fast sausage wave trains that propagate along a dense vertical current sheet to support this idea. They found that the frequency drift of the wavelet tadpoles corresponds to the drift of individual fiber bursts and they suggested the use of this information for the determination of the density profiles of the propagating magnetoacoustic wave from the fiber burst profiles measured along the radio frequency at some specific times.

%There are still a number of open questions related to the wave propagation in the current sheets. For example, the description of the propagation of the fast sausage  waves in current sheets in the gravitationally-stratified solar atmosphere can contribute to solving of the long-standing discussion about the origin of the so-called fiber (intermediate) bursts. There is still no clear evidence which types of waves are really present in radio sources of these bursts.

\cite{2017arXiv170306674J} advanced the above-mentioned studies by performing high-resolution numerical simulations of the oscillatory processes during  magnetic reconnection in a vertical, gravitationally-stratified current sheet. Development of magnetic reconnection leads to appearance of plasmoids that under the gravitational and buoyancy forces move upward or downward along the current sheet.  These plasmoids collide with each other, as well as with the underlining magnetic arcade. After the collisions the plasmoids oscillate with the periods determined by the Alfv\'en travel time within the plasmoids.  These oscillations could be responsible for the drifting pulsating structure (DPS) with distinct quasi-periodic oscillations in frequency, detected in the radio spectrum \citep{2016CEAB...40...93K}.
 
Finally, efforts have been made using kinetic theory to   model pulsations and periodicities generated by the plasma-emission mechanism of radio waves. For example, quasi-periodic generation of Langmuir waves and radio emission due to density inhomogeneities \citep{2001A&A...375..629K} and radio-emission pulsations produced via nonlinear oscillations (see \citealt{2014A&A...562A..57R}; \citealt{2016PhPl...23f2310F}; and references therein). 

%Such radio-emission modelling efforts are essential to interpreting LOFAR (LOw-Frequency ARray, \citealt{2013A&A...556A...2V}) observations and potential LOFAR detections of QPPs.

\subsection{\lq\lq{Magnetic tuning fork}\rq\rq{} oscillation driven by reconnection outflow}\label{sec : Magnetic tuning fork} 

Magnetic reconnection, the central engine of solar flares, can drive supersonic Alfv\'enic flows. Such fast reconnection outflows can be an exciter of oscillations through the collision with the ambient plasma. The oscillations excited by the reconnection outflows may have the potential to tell us about the in situ physical quantities of flares. However, the oscillation process will be highly nonlinear, because the supersonic reconnection outflows will form nonlinear waves and shocks. Therefore, direct MHD simulations are necessary for a complete understanding.

\citet{2016ApJ...823..150T} performed a set of 2D MHD simulations of a solar flare and studied oscillations excited by the reconnection outflow. Unlike previous models for quasi-periodic propagating fast-mode magnetoacoustic waves (QPFs), their model includes essential physics for solar flares such as magnetic reconnection, heat conduction, and chromospheric evaporation. From the simulations, they discovered the local oscillation above the loops filled with evaporated plasma (above-the-loop-top region) and the generation of QPFs from such oscillating region. In this section, we will introduce the physical process found in their study.

\begin{figure}
\begin{center}
\includegraphics[width=3.0in]{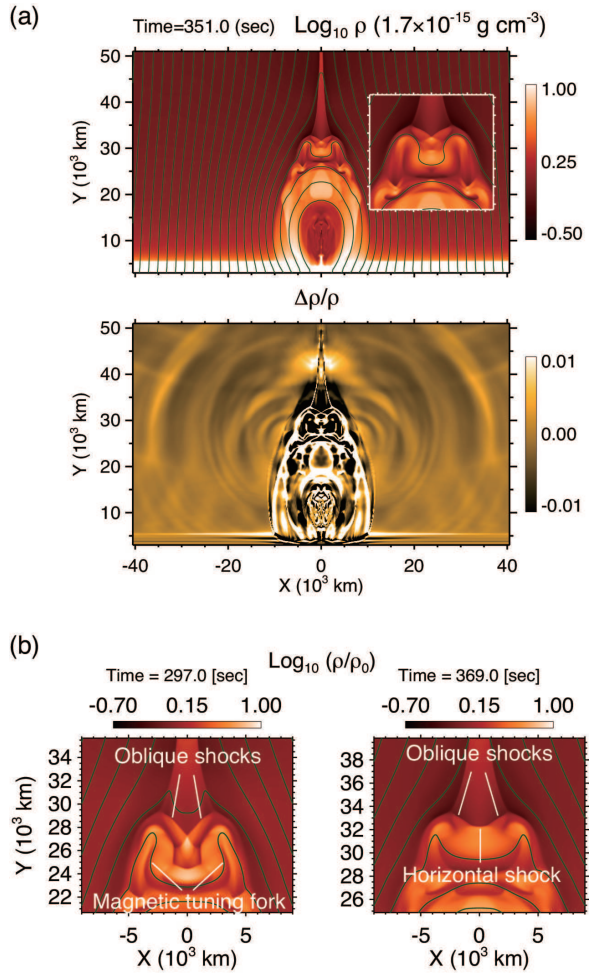}
\caption{Quasi-periodic propagating fast-mode waves (QPFs) in a simulation of \citet{2016ApJ...823..150T}. (a) Top: density map. The solid lines denote magnetic field lines. An enlarged image of the above-the-loop-top region is also displayed. Bottom: the normalized running difference of the density. (b) snapshots of the above-the-loop-top region. The colour contour indicates the density.}
\label{fig:qpf_alt}
\end{center}
\end{figure}

Figure~\ref{fig:qpf_alt}a displays snapshots of the density distribution in the site of the  simulated flare. Magnetic reconnection drives the narrow reconnection outflow. The reconnected fields pile up and form a loop system which is eventually filled with the hot dense plasma coming from the chromosphere (chromospheric evaporation). The loops filled with evaporated plasma correspond to the soft X-ray flare loops, and therefore the authors call the region above the loops ``above-the-loop-top region" (an enlarged image of this is shown in Figure~\ref{fig:qpf_alt}a). 

\citet{2016ApJ...823..150T} discovered an oscillation in the flaring region even without imposing any oscillatory perturbations. The normalised running difference image of the density ($\Delta \rho/\rho$) in Figure~\ref{fig:qpf_alt}a clearly shows the recurrent generation of isotropic waves. These waves are identified as fast-mode MHD waves. A noticeable point is that these waves are emitted from the oscillating above-the-loop-top region. Thus, the wave source of the fast-mode waves (QPFs)  or \lq\lq{quasi-periodic flows}\rq\rq{}  is very small compared to the system size (less than 10\% of the system size in this simulation).

The above-the-loop-top region is found to be full of shocks and waves (see Figure~\ref{fig:qpf_alt}a), which is different from the previous expectations based on a standard flare model and previous simulations \citep{2001ApJ...549.1160Y}. These shocks and waves are formed as a result of the collision of the supersonic reconnection outflow with the reconnected loops piled up below. In the standard flare model, a standing horizontal fast-mode shock is expected to be formed at the termination site of the outflow and is often referred to as a ``termination shock" \citep{2002A&ARv..10..313P}. However, the simulation shows that a V-shaped pattern is formed by two oblique fast-mode shocks and later by two oblique shocks and a single horizontal fast-mode shock (Figure~\ref{fig:qpf_alt}b). Moreover, the multiple termination shocks are never stationary and the structure changes drastically with time. We note that the very dynamic shocked region can be formed even in the case of steady reconnection (the localised resistivity is fixed in time and space in this study). It can be shown that this is a natural consequence of the termination of the reconnection outflow (see \citealt{2015ApJ...805..135T} and \citealt{2016ApJ...823..150T} for more details).

\begin{figure}
\begin{center}
\includegraphics[width=5.0in]{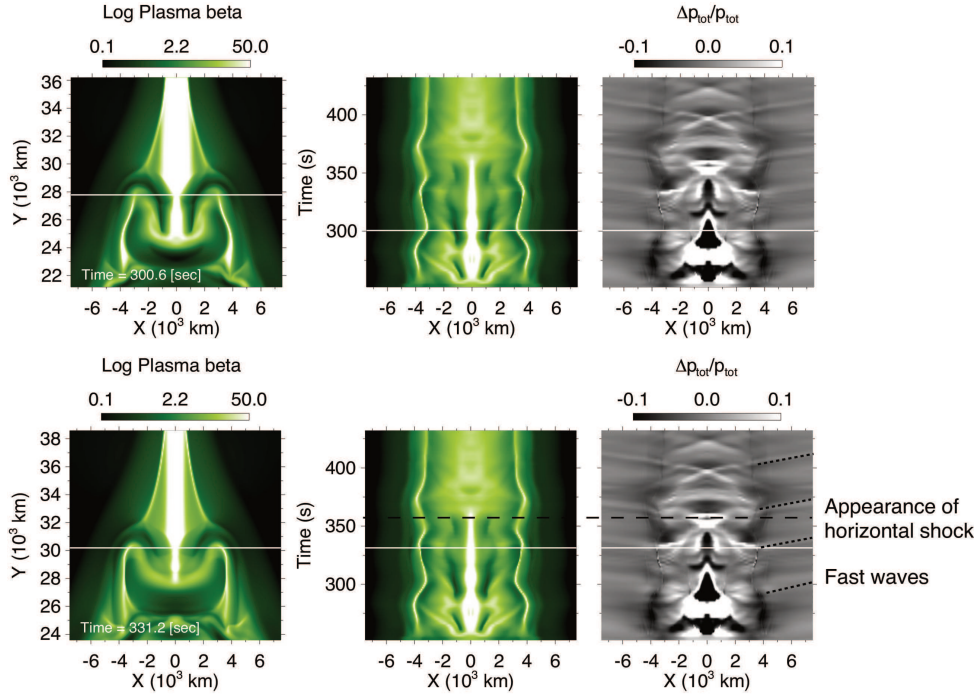}
\caption{Above-the-loop-top oscillation or magnetic tuning fork oscillation. Left: snapshots of the distribution of the plasma $\beta$ of the above-the-loop-top region. Right: time-sequenced images of the plasma $\beta$ and normalised running difference of the total pressure along the slit shown in the left panels. The horizontal lines in the time-sequenced images denote the timings of the snapshots in the left panels.}
\label{fig:tuning_fork_oscillation}
\end{center}
\end{figure}

The above-the-loop-top region shows a pair of the sharply bent magnetic field structures (see Figure~\ref{fig:qpf_alt}b). Looking at the temporal evolution, one finds that the distance between the two arms changes quasi-periodically. The oscillation is displayed in Figure~\ref{fig:tuning_fork_oscillation}. The left panels show snapshots of the plasma $\beta$ of the above-the-loop-top region. The right panels indicate time-sequenced images of plasma $\beta$ and normalised running difference of the total pressure $\Delta p_{\rm tot}/p_{\rm tot}$ obtained along the slit shown in the left panels, where the total pressure is the sum of the gas and magnetic pressures. The slit is positioned just below the V-shaped termination shocks. The figure shows that the two arms, shown as the two narrow high-$\beta$ regions at the left and right edges, are oscillating with a period of $\sim$40~s (top and bottom rows show the timings when the two arms are closed and open, respectively).

Time-sequenced images of $\Delta p_{\rm tot}/p_{\rm tot}$ in Figure~\ref{fig:tuning_fork_oscillation} indicate that outward-propagating fast-mode waves are excited quasi-periodically when the outward motion of the arms terminates. These fast-mode waves are what we have already displayed in Figure~\ref{fig:qpf_alt}a. Thus, it is clarified that the QPFs in the simulations are excited by the oscillatory motion of the above-the-loop-top region.

\begin{figure}
\begin{center}
\includegraphics[width=4.0in]{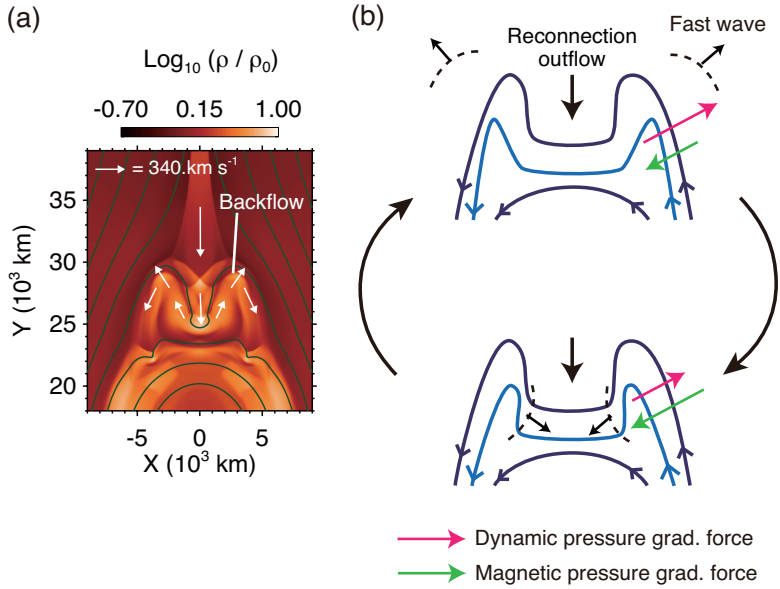}
\caption{(a) Backflow of the reconnection outflow in the above-the-loop-top region. (b) Schematic illustration of the magnetic tuning fork oscillation.}
\label{fig:tuning_fork_illust}
\end{center}
\end{figure}

What is the generation mechanism of the oscillation and QPFs? Looking at Figure~\ref{fig:tuning_fork_illust}a, one will find a fast backflow of the reconnection outflow in the small above-the-loop-top region. It was found that this backflow is the exciter of the oscillation. Figure~\ref{fig:tuning_fork_illust}b illustrates how the backflow controls the oscillation. The backflow (more exactly, the gradient of the dynamic pressure by backflow) pushes the two arms outward and compresses the magnetic field around the arms. This leads to the generation of outward-propagating fast-mode waves. Once the magnetic field there becomes strong enough to overcome the backflow, the arms start to move inward. The same process repeats and the oscillation is maintained as long as there is a strong backflow. The generation process of fast-mode waves by the backflow-driven oscillatory motion of the two arms is similar to the generation process of sound waves by an externally-driven tuning fork. For this reason, \citet{2016ApJ...823..150T} name the two arms (a pair of the sharply bent magnetic field structures) ``magnetic tuning fork". For the rest of this section,  the oscillation is  called the  ``magnetic tuning fork oscillation".

The shock structure is essential for maintaining the oscillation. The oscillation stops when a horizontal fast-mode shock appears in between the two oblique shocks (right panel of Figure~\ref{fig:tuning_fork_oscillation}b). The timing of the appearance is also indicated in Figure~\ref{fig:tuning_fork_oscillation}. The horizontal shock more significantly decelerates the reconnection outflow than oblique shocks. Therefore, the backflow should become slower after the formation of the horizontal shock. This leads to the disappearance of the oscillation. Thus, the termination shock structure is key for the maintenance of the oscillation.

The magnetic tuning fork oscillation generates not only QPFs but also the quasi-periodic oscillation of the termination shock strength. It has been argued that termination shocks could be a promising site for particle acceleration \citep{1998ApJ...495L..67T,2013PhRvL.110e1101N,2015Sci...350.1238C} and could be related to the above-the-loop-top hard X-ray source \citep{1994Natur.371..495M,2010ApJ...714.1108K,2015ApJ...799..129O}, although the detailed acceleration process in such low Mach number and high-$\beta$ shocks should be studied in more detail \citep[for recent studies about electron acceleration at such shocks, see][]{2011ApJ...742...47M,2014ApJ...794..153G}. If this is correct, it is possible that the quasi-periodic oscillation of the multiple termination shocks found in their study leads to QPPs in non-thermal emissions through the quasi-periodic variation of the efficiency of particle acceleration. In addition, the simulations showed that the oscillation of multiple termination shocks and QPFs can have a common origin. On this basis, the authors suggest a new picture in which QPFs and QPPs in the non-thermal emissions have a common origin. Thermal emissions may also respond to the variation of the efficiency of acceleration through the thermalisation of non-thermal particles, showing QPPs.

What underlying physics determines the periodicity? Since the magnetic tuning fork oscillation is controlled by the backflow of the reconnection outflow in the above-the-loop-top region, we expect that the oscillation period $P$ is proportional to the timescale determined by the backflow in this region: $P\propto w/v_{\rm bf}$, where $w$ is the size of the above-the-loop-top region and $v_{\rm bf}$ is the backflow speed. This expectation is supported by a set of numerical simulations with different magnetic field strength $B$ (equivalently, different plasma  $\beta$ in their study). Therefore, the problem is to clarify what determines the dependencies of the backflow speed $v_{\rm bf}$ and the size $w$.

The simulations show that the backflow speed is of the order of the Alfv\'en speed, as expected. This indicates that $v_{\rm bf}\propto B\propto \beta^{-1/2}$. The dependence of the size $w$ could be determined by the balance between the mass flux of the reconnection outflow and the mass flux carried horizontally by the backflow. Considering the fundamentals of reconnection and shock theories, gives  $w \propto \beta^{4/7}L^{9/7}$ (see \citealt{2016ApJ...823..150T} for  details). From these results, we obtain the following scaling relation:
\begin{align}
P\propto \frac{w}{v_{\rm bf}}\propto \beta^{15/14}L^{9/7}\propto B^{-2.1}\label{eq:scaling_period}
\end{align}
Numerical results are consistent with this predicted scaling relation. This further supports the fact that the oscillation is controlled by the backflow of the reconnection outflow.

It is noteworthy that the period is sensitive to the magnetic field strength (see equation \ref{eq:scaling_period}). In the typical case of a flare with the size of 60~Mm and plasma $\beta$ of 0.08, the oscillation has a period of $\sim$40~s. Since the real coronal field strength will vary from flare to flare, typically from a few G to several tens G or larger, the magnetic tuning fork oscillation can create a large range of periods. This is consistent with the fact that QPFs and QPPs have a wide range of periods. Attention must be paid when one directly applies the scaling relation to observations, because the effects of the 3D global magnetic field structure and time-dependent magnetic reconnection, which are not considered in the study, can make the process more complex. Investigating these effects will be important for advancing the theory.

{{Recently, \citet{2017arXiv170905234T} performed a series high-resolution 2D simulations of magnetic reconnection which occurs below an erupting CME. They showed that the oscillation in the above-the-loop-top region operates even in the case of plasmoid-dominated reconnection, although the oscillation tends to be asymmetric and the dynamics in the current sheet becomes turbulent in this case. Since the reconnection jets are bi-directional, two oscillation periods from the two termination shock regions emerge in their simulations. \citet{2017arXiv170905234T} presents an observational example of a flare associated with a CME which shows two distinct periods in the QPPs. Another important finding is that the magnetic tuning fork oscillation at the bottom of a CME can cause quasi-periodic oscillations of the CME. Since such oscillations of CMEs have been reported \citep[e.g.][]{2001ApJ...562.1045K}, it will be interesting to statistically investigate the relationship between oscillations of CMEs and QPPs in emissions.}}

The above-the-loop-top region is very small, and therefore resolving a detailed structure of this region is observationally challenging. Nevertheless, confirming the key features of the magnetic tuning fork oscillation from observations is necessary to validate this mechanism. It is important to affirm the presence of the fast backflow from spectroscopic observations. The oscillation can lead to a quasi-periodic pulsations of emissions from the above-the-loop-top region. Therefore, investigating the temporal evolution of emissions from this region may allow us to infer whether the magnetic tuning fork oscillation actually operates. If the pulsations in the above-the-loop-top region and QPFs have a similar period, then this can be strong supporting evidence.

{{Direct identification of the magnetic tuning fork oscillation is very challenging. Therefore, it is necessary to accumulate pieces of indirect supporting evidence. For this aim, the following (and probably more) considerations will be useful. Bursts in non-thermal emissions, which seem to occur quasi-periodically in some cases, are sometimes attributed to the electron acceleration associated with plasmoid dynamics \citep{2006Natur.443..553D,2015ApJ...799..126N,2016ApJ...828..103T}. One has to check if QPPs are accompanied by plasmoid ejections/dark downflows from coronal images and/or dynamic radio spectra (since plasmoids may be seen as drifting pulsating signatures in dynamic radio spectra, \citealt{2004A&A...417..325K}). The quasi-periodic change of the  magnetic reconnection rate is also a possible cause for QPPs, and plasmoid-dominated reconnection and oscillatory reconnection have been discussed in this context \citep{2000A&A...360..715K,2009A&A...493..227M}. If the change in reconnection rate is the main cause of QPPs, then QPPs in emissions and the reconnection rate estimated from e.g. flare ribbon separation \citep[e.g.][]{2005ApJ...632.1184I} will correlate well. These do not necessarily correlate in the case of the magnetic tuning fork oscillation (of course this is also true for many other cases). We need to examine if standing MHD waves in flare loops \citep[e.g.][]{1999ApJ...513..516C,2004A&A...414L..25N} are irrelevant to the observed oscillations or not. This may be done by tracking the motion of reconnected field lines or hard X-ray loop top source \citep[][]{2006ApJ...644L..97L} and from Doppler observations \citep{2005ApJ...620L..67M}. We are aware that the amplitude of the oscillation in \citet{2006ApJ...644L..97L} is estimated to be only approximately 300~km ($\sim$0.4$^{\prime\prime}$) which is much smaller than the spatial resolution of {\it RHESSI}, $\sim$7$^{\prime\prime}$. Therefore, there needs to be caution with regards to the interpretation of such a result.}}

%As discussed in \citet{2017arXiv170905234T}, the oscillation in the termination shock region at the bottom of the CME can lead to oscillations of the CME. Investigating the relationship between oscillations of CMEs and QPPs in emissions will be interesting.

%The above-the-loop-top region is very small, and therefore resolving a detailed structure of this region is observationally challenging. Nevertheless, confirming the key features of the magnetic tuning fork oscillation from observations is necessary to validate this mechanism. It is important to affirm the presence of the fast backflow from spectroscopic observations. The oscillation can lead to a quasi-periodic pulsations of emissions from the above-the-loop-top region. Therefore, investigating the temporal evolution of emissions from this region may allow us to infer whether the magnetic tuning fork oscillation actually operates. If the pulsations in the above-the-loop-top region and QPFs have a similar period, then this can be strong supporting evidence. To clarify that the oscillating region is localised in the above-the-loop-top region, one has to check if other processes such as periodic plasmoid ejections \citep[e.g.][]{2000A&A...360..715K} and standing MHD waves in the flare loops \citep[e.g.][]{2004A&A...414L..25N} are irrelevant to the observed oscillations or not.

\subsection{Wave-driven reconnection in the Taylor problem} \label{sec : wave-driven reconnection in the Taylor problem}

As mentioned in \S\ref{sec:OscillatoryReconnection}, the seminal reconnection models of Sweet--Parker (\citealt{1957JGR....62..509P}; \citealt{1958IAUS....6..123S}) and Petschek  \citep{1964NASSP..50..425P} are steady-state models. Apart from the obvious problem of applying steady-state theory to dynamic flaring events, there is a subtler issue here: steady-state models can only provide one timescale, that of steady reconnection; proportional to $S^{1/2}$ for Sweet-Parker, and to $\ln{S}$ for Petschek \citep{2004ARA&A..42..365B}. However, flares require a growth rate that is not only fast, but also exhibits a sudden increase in its time derivative. This is referred to as the \lq{trigger problem}\rq{}; the magnetic topology evolves slowly for an extended duration, only to then undergo a sudden change over a shorter timescale. Sweet--Parker and Petschek cannot account for the time evolution of the reconnection rate. Instead, time-dependent reconnection rates are referred to as impulsive or bursty. In addition, magnetic reconnection can be broadly classified into two types: {\emph{free}} or {\emph{spontaneous}}, i.e. caused by an intrinsic instability which taps into the magnetic free energy stored within the equilibrium topology, or {\emph{forced}}, which is driven by perturbations (e.g. from a boundary) that induce a change in connectivity to an equilibrium or a sudden increase in the anomalous resistivity in the reconnection site.

An interesting forced, impulsive mechanism has been investigated with regards to the so-called \lq{Taylor}\rq{} problem, which was proposed by J. B. Taylor in a private communication to \citet{1985PhFl...28.2412H} who then carried out the analysis themselves.  \citet{1985PhFl...28.2412H} considered a slab plasma equilibrium that is suddenly subjected to imposed, small-amplitude boundary perturbations such as to drive magnetic reconnection at the centre of the slab. The  so-called \lq{Taylor}\rq{} problem has then been developed by several authors, including \citet{1992PhFlB...4.1795W}, \citet{1996PhPl....3.2129W} and \citet{2003PhPl...10.2304F}, who all consider slight variations on the initial and boundary conditions of the fundamental system, and \citet{2015PhPl...22d2109C} who extended the theory into the plasmoid-unstable regime. A self-consistent solution for the continuous plasma-heating was derived by \citet{1999PhPl....6.2897V}  within  a similar system undergoing forced external driving. The authors found that the plasma-heating rate displayed a relaxation-type dependency on the driving frequency, leading to a discussion whether forced magnetic reconnection can be interpreted as an Alfv\'en resonance with zero frequency \citep{1999JPlPh..62..345U} or not \citep{2000PhPl....7.3808V}.

%ote aims to clarify the correspondence between forced magnetic reconnection and Alfv\UTF{00E9}n resonances in a plasma with a sheared magnetic field subjected to continuous external driving. It is shown how a transition from one regime to another occurs, and what implications this has on the magnetic energy dissipation rate.  {2000PhPl....7.3808V}

% All the  Taylor-problem models are restricted to 2D nonlinear MHD simulations.

As an illustrative example, here we adopt the conditions of \citet{2003PhPl...10.4284F}. Initial conditions are chosen such that ${\bf{B}} = \left(0, B_0 x , B_T  \right)$, where $B_0$ and $B_T$ are constants, and ${\bf{A}}= \left[0,0,\phi(x,y,t)\right]$ is the vector potential. The plasma is stable to tearing modes. The plasma is bounded by perfectly conducting walls (located at $x=\pm a$) and is periodic in the $y-$direction with periodicity length $L$. The conducting walls are subject to displacement $\Xi(t)\cos{\left(k y\right)}$ at $x=a$ where $k=2\pi / L$. An equal and opposite displacement is applied at $x=-a$. Here $\Xi(t)\propto 1- \exp{\left( {t/\tau} \right) }-\left(t/\tau  \right)\,\exp{\left( {-t/\tau} \right) }$ is a ramp-up term.  At early times, the plasma builds up a concentration of current (surface current) along $x=0$ (the resonant surface). Subsequently, reconnection of flux across the resonant surface occurs, forming a chain of magnetic islands. As the reconnection proceeds, the surface current decreases and the plasma tends towards equilibrium. The process evolves through multiple stages in the reconnection process, labelled $A$, $B$, $C$, $D$ and $E$ by  \citet{1985PhFl...28.2412H}. Phases $A$ to $C$ are governed by linear boundary-layer physics, phase $D$ corresponds to Sweet-Parker reconnection (under certain circumstances, see \citealt{1992PhFlB...4.1795W}) and  phase $E$ corresponds to the nonlinear magnetic island dynamics of \citet{1973PhFl...16.1903R}. 

\citet{1985PhFl...28.2412H}, \citet{1992PhFlB...4.1795W}, \citet{1996PhPl....3.2129W} and \citet{2003PhPl...10.2304F} investigated the Taylor problem using Laplace Transforms and solved the resulting equations via asymptotic matching, equivalent to a boundary-layer problem. Essentially, there is an assumption that the plasma is divisible into two regions: an {\emph{inner}} region (non-ideal, time-dependent, narrow) around the $x=0$ resonant surface and an {\emph{outer}} region (ideal, steady) consisting of the rest of the plasma \citep[see, e.g.][]{2004ARA&A..42..365B}. The inner and outer solutions are then matched asymptotically; referred to conventional asymptotic theory. \citet{2003PhPl...10.4284F} developed an improved Laplace Transform approach, which does not involve asymptotic matching,  to investigate the early time response of the plasma; which is missed by the  boundary-layer approach. 

%From Ohm's law:
%\begin{equation}
%\frac{d \phi_0} {dt} = \eta J(t),
%\end{equation}
%where $\phi_0(t)$ is the reconnected magnetic flux and $J(t)$ is the amplitude of the current sheet driven at the resonant surface (as well as proportional to the rate of magnetic reconnection).

\begin{figure}
\begin{center}
\includegraphics[width=2.5in]{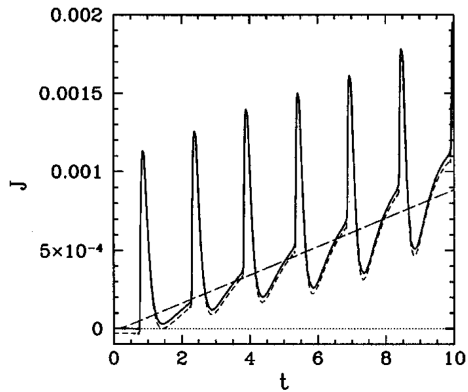}
\caption{Evolution of the magnetic reconnection rate, $J(t)$. Solid curve shows numerical solution generated by the FLASH code. Long-dashed curve shows the solution produced by conventional asymptotic matching theory. The short-dashed curve shows the solution generated by the improved Laplace Transform approach (and has been shifted downwards slightly to make it more visible). Reproduced from \citet{2003PhPl...10.4284F}.}
\label{figure_Fitzpatrick_et_al_2003}
\end{center}
\end{figure}

If the wall perturbation is switched on slowly compared to the Alfv\'en time, then the plasma response eventually asymptotes to that predicted by conventional asymptotic theory. However, at early times there is a compressible Alfv\'en wave driven contribution to the reconnection rate which leads a significant increase in the reconnection rate. If the wall perturbation is switched on rapidly compared to the Alfv\'en time then strongly localised compressible Alfv\'en wave pulses are generated which bounce back and forth between the walls several times. Each time these wave-pulses cross the resonant surface, they generate a transient surge in the reconnection rate. The maximum pulse-driven reconnection rate is much larger than that from conventional asymptotic theory. The evolution of the current density is dominated by a series of spikes in the reconnection rate. This can be seen in Figure \ref{figure_Fitzpatrick_et_al_2003}.

\citet{2003PhPl...10.4284F} derive an expression for the evolving current (see their equation 30) that involves integrating over the Bromwich contour. The numerator of the integrand contains multiple poles which correspond to those of the function $\partial Y / \partial x|_{x=0}$ where $Y(x,g)$ is a solution of the linearised, Laplace-transformed MHD equations governing the system. The poles can be written $g=\pm i \omega_n$ where $\omega_n$ (which are real) represents the oscillation frequencies associated with the natural Alfv\'enic modes of oscillation of the plasma. The plasma response emanating from these poles can be thought of as due to compressible Alfv\'en waves  excited by the sudden imposition of the wall perturbation \citep{2003PhPl...10.4284F}. 

%{{Note that \citet{2003PhPl...10.4284F} uses the terminology {\it{compressible}} and {\it{compressional}} Alfv\'en wave. When pressure gradients are included, the compressional Alfv\'en wave becomes a fast magnetoacoustic wave (see $\S4.3.2$ of \citealt{2014masu.book.....P}) and the model of  \citet{2003PhPl...10.4284F} does involve pressure gradients. In the solar literature, compressional Alfv\'en  waves  are usually called fast magnetoacoustic waves  even in the case of pressure being neglected (so as to avoid confusion with the shear  Alfv\'en wave or, in cylindrical geometry, torsional Alfv\'en wave which is sustained soley by magnetic tension). Thus, in accordance with the solar literature, for the rest of this section we use the terminology {\it{fast magnetoacoustic}}.}}

{{
Note that \citet{2003PhPl...10.4284F} uses the terminology {\it{compressible}} and {\it{compressional}} Alfv\'en wave. In the solar literature, compressional (or, actually, compressive) Alfv\'en waves are usually called fast magnetoacoustic waves, even in the case of gas pressure being neglected and the wave is driven by the magnetic pressure gradient (so as to avoid confusion with the incompressive shear Alfv\'en wave or, in cylindrical geometry, torsional Alfv\'en wave which are both sustained solely by magnetic tension). Thus, in accordance with the solar literature, for the rest of this section we use the terminology fast magnetoacoustic.

}}

%(these pulses can be  described as fast magnetoacoustic pulses; when pressure gradients are included, the compressional Alfv\'en wave becomes a fast magnetoacoustic wave; 

Simulations by \citet{2003PhPl...10.4284F} show that this manifests such that the sudden switch-on of the wall perturbation generates two strongly localised pulses which propagate towards the resonant surface, pass through one another and reflect off the walls. The two pulses then subsequently bounce back and forth several times. The arrival time of the pulses at $x=0$ correlates with the spikes in the reconnection rate.  The authors conclude that the strong spikes in $J(t)$, {{ i.e. the amplitude of the current sheet driven at the resonant surface,  }} represent magnetic reconnection driven by {{fast magnetoacoustic  waves}}  which are excited by the sudden onset of the wall perturbation. The physical mechanism as to why the reconnection rate increases sharply as the {{fast magnetoacoustic wave}}  pulse transits the resonant layer is not reported, although we may speculate that this is related to an increase in gradients of ${\bf{B}}$ and therefore $\nabla \times {\bf{B}}$ and $J(t)$ as the wave passes through the resonant layer.

% need to check Fitzpatrick section  .... 

%{{Increasing the central pressure increases the propagation speed (these pulses can be  described as fast magnetoacoustic pulses - when pressure gradients are included, the compressional Alfv\'en wave becomes a fast magnetoacoustic wave;  see $\S4.3.2$ of \citealt{1982soma.book.....P}) and hence decreases the period between the spikes in $J(t)$. The period is governed by the fast magnetoacoustic speed and the distance between the walls $2a$, i.e. the travel time.  Since this is a theoretical model solely, all periods can be obtained by tuning the fast magnetoacoustic speed and  $a$.}}

%The period is governed by the fast magnetoacoustic speed and the distance between the walls $2a$, i.e. the travel time.  For example, increasing the central pressure increases the (fast) propagation speed and hence decreases the period between the spikes in $J(t)$. Since this is a theoretical model solely, all periods can be obtained by tuning the fast magnetoacoustic speed and  $a$.

{{

The period is governed by the fast magnetoacoustic speed  and the distance between the walls $2a$, i.e. the travel time.  For example, increasing the central pressure increases the (fast magnetoacoustic) propagation speed and hence decreases the period between the spikes in $J(t)$. Since this is a theoretical model solely, a broad range of periods, e.g. from seconds to minutes, can be obtained by tuning the fast magnetoacoustic speed and  $a$.

}}
%The period is governed by the fast magnetoacoustic speed  (these pulses can be  described as fast magnetoacoustic pulses; when pressure gradients are included, the compressional Alfv\'en wave becomes a fast magnetoacoustic wave;  see $\S4.3.2$ of \citealt{1982soma.book.....P}) and the distance between the walls $2a$, i.e. the travel time.  For example, increasing the central pressure increases the (fast) propagation speed and hence decreases the period between the spikes in $J(t)$. Since this is a theoretical model solely, a broad range of periods, e.g. from seconds to minutes, can be obtained by tuning the fast magnetoacoustic speed and  $a$.

%Since this is a theoretical model solely, all periods can be obtained by tuning the fast magnetoacoustic speed and  $a$.

We note that the pulses only remain coherent over several transits where $k \ll 1$, equivalent to $L \gg a$. This is equivalent to the wavelength of the wall perturbation $\propto k^{-1}$ being much greater than the wall separation $\propto a$, which places limitations on the applicability of the model.

%All the  Taylor problem models considered so far are restricted to 2D nonlinear MHD and so a natural progression would be to extend the modelling to three-dimensions. In addition, the periodicity of the model is based on reflections between opposing solid boundaries and thus its applicability to {{solar and stellar}} atmospheres is unclear (whereas its applicability to, e.g., tokamaks is more obvious). In the solar atmosphere, the requirement for solid boundaries is usually achieved via, say, bouncing repeatedly between two footpoints in a wave-guiding coronal loop (this is of course not the only scenario{{; one can imagine a reconnection region might be bounded by strong magnetic field, which would act as walls enveloping a cavity). As an example for stellar objects, \citet{2008MNRAS.384.1355G} reported on a periodicity in the  flaring rate of  binary star YY Gem and proposed that magnetic reconnection (responsible for the flaring) is modulated by fast-mode magnetoacoustic waves which are trapped between the surfaces of the two stars, so that the reconnection rate presents a periodic behaviour.}}  However, in such systems, the compressional Alfv\'en waves would take a different form and one would require a resonant layer (as defined in the Taylor problem, not resonant absorption) to be located at the loop apex. It is unclear how this could be achieved topologically, let alone the wave generation aspect. However, from  a forced, impulsive reconnection mechanism point-of-view, these papers are seminal.

All the  Taylor problem models considered so far are restricted to 2D nonlinear MHD and so a natural progression would be to extend the modelling to three-dimensions. In addition, the periodicity of the model is based on reflections between opposing solid boundaries and thus its applicability to {{solar and stellar}} atmospheres is unclear (whereas its applicability to, e.g., tokamaks is more obvious). In the solar atmosphere, the requirement for solid boundaries is usually achieved via, say, bouncing repeatedly between two footpoints in a wave-guiding coronal loop.{{ However, in such systems, the fast magnetoacoustic  waves would take a different form and one would require a resonant layer (as defined in the Taylor problem, not resonant absorption) to be located at the loop apex. It is unclear how this could be achieved topologically, let alone the wave generation aspect.}} This is of course not the only scenario{{; one can imagine a reconnection region might be bounded by strong magnetic field, which would act as walls enveloping a cavity). As an example for stellar objects, \citet{2008MNRAS.384.1355G} reported on a periodicity in the  flaring rate of  binary star YY Gem and proposed that magnetic reconnection (responsible for the flaring) is modulated by fast magnetoacoustic waves which are trapped between the surfaces of the two stars, so that the reconnection rate presents a periodic behaviour.  Regardless of specific applicability, }}from  a forced, impulsive reconnection mechanism point-of-view, these papers are seminal.

%In the interbinary space, the surfaces of the two stars can serve as two walls, and indeed the waves bouncing back and forth would modulate the stellar flares as demonstrated by MHD numerical simulation of Gao et al. (2008, MNRAS, 384, 1355);

%The binary YY Gem shows many interesting properties, one of which is the periodicity in its flaring rate. The period, which is about 48 +/- 3 min, was ever interpreted in terms of the oscillation of a filament. In this paper, we propose a new model to explain this phenomenon by means of 2.5D magnetohydrodynamic (MHD) numerical simulations. It is found that magnetic reconnection is induced as the coronal loops rooted on both stars inflate and approach each other, which is driven by the differential stellar rotation. The magnetic reconnection is modulated by fast-mode magnetoacoustic waves which are trapped between the surfaces of the two stars, so that the reconnection rate presents a periodic behaviour. With the typical parameters for the binary system, the observed period can be reproduced. We also derive an empirical formula to relate the period of the flaring rate to the coronal temperature and density, as well as the magnetic field. 

\subsection{Two loop coalescence}\label{sec : Two loop coalescence}     

\begin{figure}
\begin{center}
\includegraphics[width=4.0in]{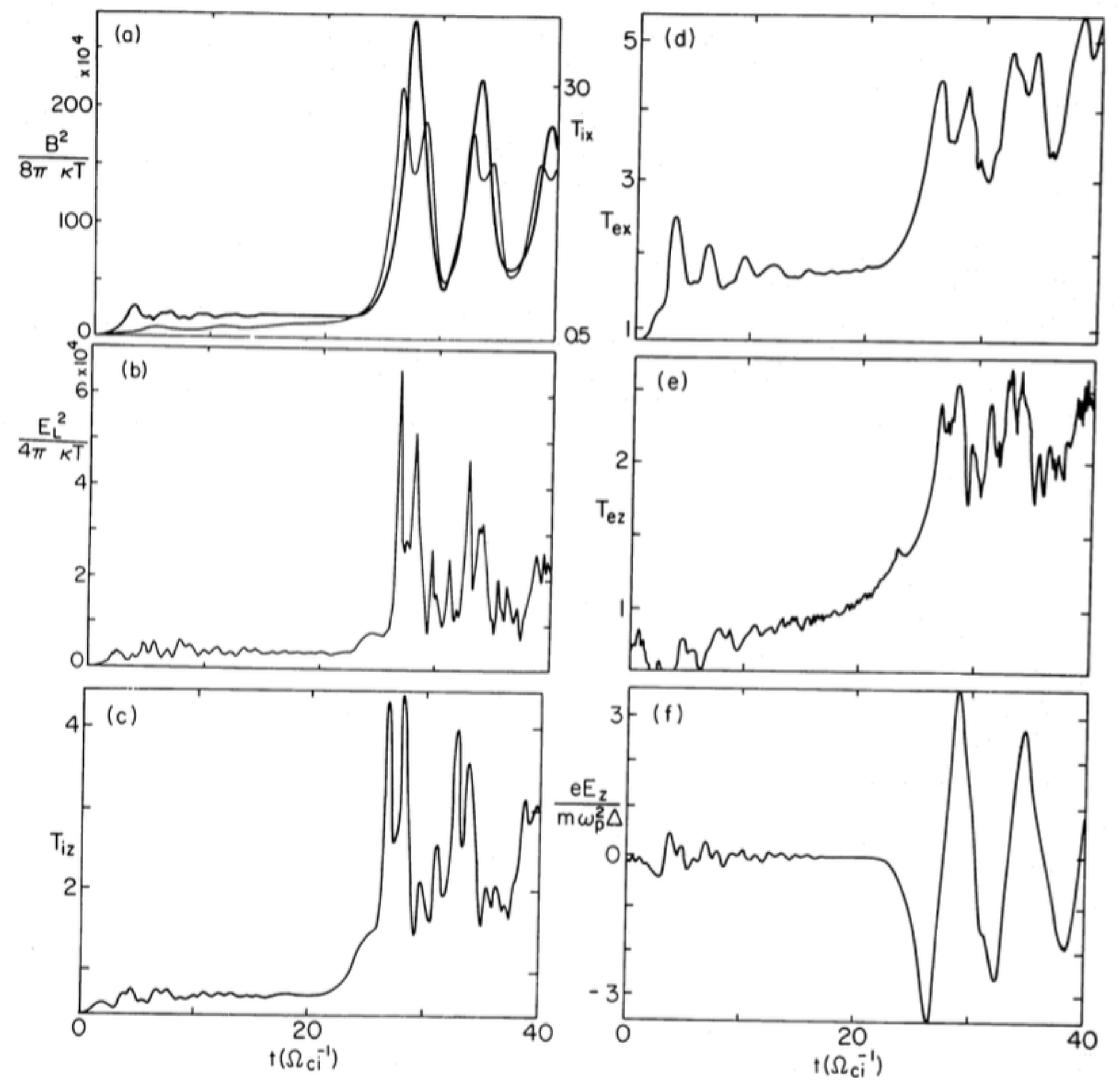}
\caption{Temporal profiles of particle and field quantities for the coalescence process. (a) The thick line represents the magnetic energy, the thin one the ion temperature in the $x$-direction. (b) Electrostatic field energy in time. (c) Ion temperature in the $z$-direction. (d) Electron temperature in the $x$-direction. (e) Electron temperature in the $z$-direction. (f) Inductive electric field ($E_z$). From \citet{1987ApJ...321.1031T}.}
\label{fig:tajima_fig7}
\end{center}
\end{figure}

It has been widely believed that a current sheet with a high ($> 10^4$) Lundquist number is subject to fragmentation \citep{2016ASSL..427..373S}. This instability is called the plasmoid instability and leads to the formation of plasmoids in a current sheet \citep{2001ApJ...551..312T,2007PhPl...14j0703L,2009PhPl...16k2102B}, where plasmoids are magnetically-confined plasma. Plasmoids in the current sheet will contain the electric current in almost the same direction with each other, because the plasmoid formation results in the {{discretization}} of the electric current in the current sheet. Therefore an attracting Lorentz force will operate between these plasmoids, which will lead to the coalescence of plasmoids. This process, known as the coalescence instability \citep{1977PhFl...20...72F,1979PhFl...22.2140P}, is considered as a key process for the bursty, impulsive energy release during solar flares \citep{1980PhRvL..44.1069B,1983PhFl...26.3332B}. The coalescence of plasmoids were actually observed \citep{2012ApJ...745L...6T}.

Previous nonlinear simulations revealed that the coalescing plasmoids show quasi-periodic oscillations in fields and other particle quantities \citep{1979PhFl...22.2140P,1987ApJ...321.1031T}. Since the particle acceleration is expected during the coalescence \citep{1982ApJ...258L..45T,2010ApJ...714..915O,2011ApJ...733..107K}, QPPs seen in emissions originated from non-thermal, high-energy particles could be caused by the oscillations associated with the coalescence instability (an observational example is given by \citealt{2016ApJ...828..103T}). On this basis, \citet{1987ApJ...321.1031T} investigated the characteristics of the oscillations in detail and compared their results with observations. As a representative work on the oscillations caused by the coalescence of plasmoids, we here briefly introduce their study in the following. 
This work has been generalised by \citet{2016PhRvE..93e3205K}, which will be also mentioned later.

\citet{1982ApJ...258L..45T,1987ApJ...321.1031T} was motivated particularly by the observations of the (Seven-Sisters) flare on 7 June 1980 that showed seven successive pulses with a quasi-periodicity of $\sim 8$~s \citep{1983Natur.305..292N}. This flare was observed in hard X-ray, gamma-ray, and microwave emissions. All of the pulses in these bands were almost synchronous within $\pm 2.2$~s and have a similar shape. The observations suggest a quasi-periodic acceleration of both electrons and ions. An interesting feature of the pulsation is that a few of the pulses of microwaves at 17~GHz showed double subpeaks. It seems that the first subpeak coincides with the peak of the corresponding hard X-ray, while the second subpeak coincides with the peak of the corresponding gamma-ray pulse. To understand the observational characteristics, the authors investigated the coalescence of plasmoids numerically and theoretically in detail. They performed both particle and MHD simulations in 2.5D (two spatial dimensions $x,y$ and three velocity and field dimensions), but here we only focus on the results of the particle simulations to see the relation between the oscillation and particle acceleration. The initial setup of their typical simulations {{assumes}} two plasmoids that are attracted by the Lorentz force by each other. A uniform external magnetic field, $B_z$, is applied. A more detailed explanation for the setup is given in \citet{1982PhFl...25..784L}.

Figure~\ref{fig:tajima_fig7} displays the temporal evolution of particle and field quantities during the coalescence process. Looking at the magnetic energy (thick line in Figure~\ref{fig:tajima_fig7}a), one sees three peaks clearly. Corresponding {{to}} these peaks, the electrostatic field $E_L$ and ion and electron temperatures show double subpeaks. Strong particle acceleration in the $z$-direction occurs at the subpeaks of $E_L$ through the $\vec{E}_L\times \vec{B}$ acceleration.

\begin{figure}
\begin{center}
\includegraphics[width=5.0in]{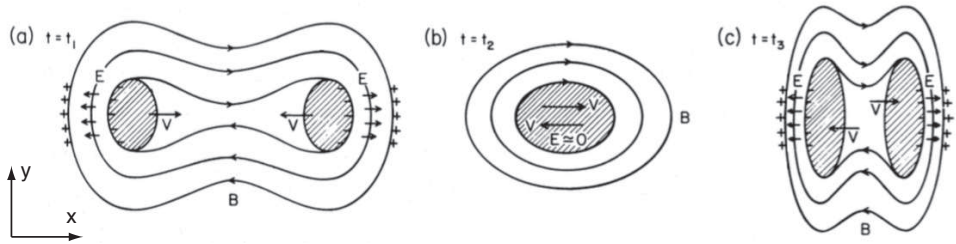}
\caption{Schematic sequence of snapshots of the plasma and electric and magnetic fields during the coalescence process. Note that the direction of the electric field is defined in the opposite way as usual. From \citet{1987ApJ...321.1031T}.}
\label{fig:tajima_fig8}
\end{center}
\end{figure}

This oscillation process is schematically described in Figure~\ref{fig:tajima_fig8}. When the coalescence starts, a strong acceleration of ions by the Lorentz force takes place. At $t=t_1$, the acceleration reaches a maximum, leading to a strong compression at the far sides of the two plasma blobs. This compression causes the first temperature peak. In addition, the difference in inertia between ions and electrons results in charge separation at the compressed regions. The charge separation generates the electrostatic field $\vec{E}_L$. The $\vec{E}_L \times \vec{B}$ acceleration, together with the magnetic acceleration, produces high energy particles in the $z$-direction. At time $t=t_2$ when two plasma blobs totally merge, the inductive electric field ($\vec{v}\times \vec{B}$) vanishes. The direction of the acceleration of electrons and ions along the $z$-direction is reversed at this moment. At time $t=t_3$, the overshooting plasma blob motions result in the generation of the reverse electrostatic field. The temperature again reaches a maximum. This is the mechanism's interpretation of the formation of the double peaks in the electrostatic field and temperature.

\citet{1987ApJ...321.1031T} developed a theoretical model of the coalescence of two plasmoids. They assume that $\partial /\partial x \gg \partial /\partial y, \partial /\partial z,$ where $x$ is the direction of coalescence so that the dynamics of the coalescence is treated as a one-dimensional problem. To separately deal with the dynamics of ions and electrons, they start from the two-fluid ideal MHD equations.

Since no specific scale length appears in this one-dimensional coalescence process, we expect the presence of a self-similar solution for this problem. They introduced scale factors $a(t)$ and $b(t)$ in the following way to look for self-similar solutions:
\begin{align}
v_{ex}=\frac{\dot{a}}{a}x \quad \quad v_{ix}=\frac{\dot{b}}{b}x
\end{align}
where a dot represents the time derivatives and $v_{ex}$ and $v_{ix}$ are the electron and ion velocities in the $x$-direction, respectively. An ansatz imposed here is that the velocities are linear in $x$. Further assuming quasi-neutrality ($n_i=n_e$), we get $a=b$. As a result, it is found that a self-similar solution for the fields and particle quantities can be written as a function of the scale factor $a(t)$. 

\begin{figure}
\begin{center}
\includegraphics[width=4.0in]{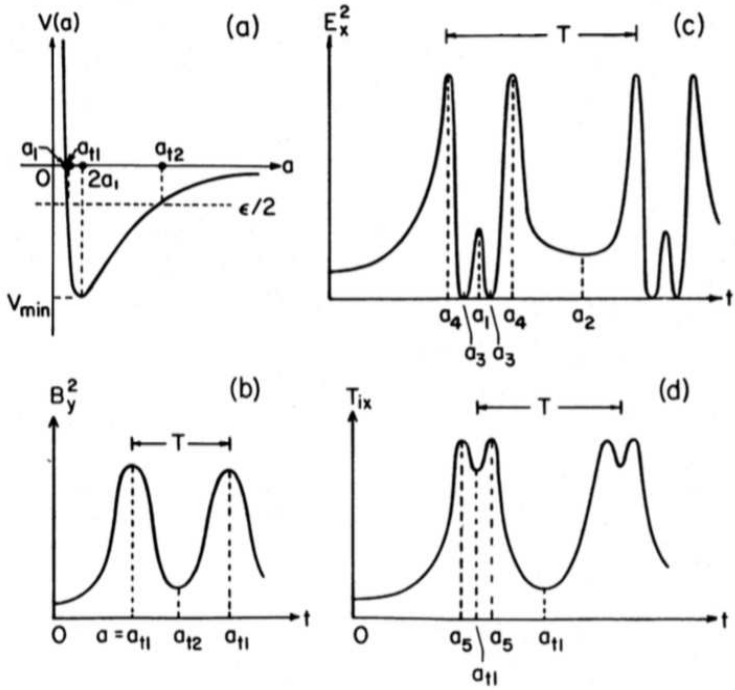}
\caption{Schematic behaviour of the explosive collapse. (a) The Sagdeev potential for the scale factor of the explosive coalescence. (b) The temporary behaviour of the magnetic field energy constructed from the Sagdeev potential. (c) The temporal behaviour of the electrostatic field energy constructed from the Sagdeev potential. (d) The temporal behaviour of the ion temperature in the $x$-direction (in the direction of coalescence) constructed from the Sagdeev potential. From \citet{1987ApJ...321.1031T}.}
\label{fig:tajima_fig12}
\end{center}
\end{figure}

The equation that governs the temporal evolution of the scale factor $a(t)$ can be written as follows:
\begin{align}
\ddot{a}=-\frac{\partial V(a)}{\partial a},
\end{align}
where the effective potential $V(a)$ has essentially the same functional form as that of the effective potential for the gravitational force (see Figure~\ref{fig:tajima_fig12}a). Thus, this means that the scale factor $a$ (and therefore the other quantities) oscillates within a finite range. The minimum oscillation period $P_{\rm min}$ is estimated as: 
\begin{align}
P_{\rm min} = 2\pi \frac{{C_\mathrm{s}}^3}{v_A^4}\lambda \simeq 2~s \left( \frac{\beta}{0.1}\right)^{3/2} \left( \frac{\lambda}{10^4~\rm km}\right)\left( \frac{v_A}{10^3~\rm km~s^{-1}}\right)^{-1}\label{eq:period_coalescence} 
\end{align}
where $C_\mathrm{s}$ and $v_A$ are the sound and Alfv\'en speeds, respectively, and $\lambda$ is a characteristic scale length of the magnetic field in the interaction region (typically the size of plasmoids). One can see that the period increases as the plasma $\beta$ increases. More detailed investigation implies that the period also depends on the magnetic twist of the plasmoids (namely, the ratio of the toroidal to poloidal magnetic fields) and the colliding velocity of the plasmoids.

Figure~\ref{fig:tajima_fig12} shows a schematic temporal behaviour of (b) the electrostatic field $E_x^2$ ($E_L^2$), (c) magnetic energy $B_y^2$, and (d) ion temperature $T_{ix}$. The double subpeak is prominent in the $E_x^2$ and $T_{ix}$ profiles, although $E_x^2$ shows another small peak. The triple-peak profile will be a double-peak profile when the plasma $\beta$ is sufficiently small. 

From numerical and analytical investigations, it is found that the coalescence results in the quasi-periodic particle acceleration as seen in the observations of a flare on 7 June 1980. The double-peak profiles of the electrostatic field and ion and electron temperatures are likely to be relevant to the double subpeaks in emissions. If one looks at the electron energy spectrum, it can be fitted by a double power law with a break. This is consistent with the observations \citep{1983ApJ...271..376K}.

The theoretical discussion of \citet{1987ApJ...321.1031T} is based on the quasi-neutrality assumption, which is valid only at the non-kinetic scale (current sheet thickness is larger than the Debye length --- We are aware that this assumption may not be consistent with the interpretation where the charge separation at a kinetic scale is discussed, see Figure~\ref{fig:tajima_fig8}). \citet{2016PhRvE..93e3205K} generalised the discussion to cover both the kinetic and non-kinetic scales. When the current sheet is thicker than the kinetic scale, the electrostatic field $E_{x}$ is produced by the mechanical effects (coming from the momentum equations). This $E_{x}$ appears from the equation of motion for ions. On the other hand, when the current sheet is comparable to or thinner than the kinetic scale, $E_{x}$ is mainly produced by the charge separation. \citet{2016PhRvE..93e3205K} found that the double-peak structure in $E_{x}$ will be a general feature and can be found in both the kinetic and non-kinetic scales, although the shape of the profile is different from that in \citet{1987ApJ...321.1031T}. In addition to this, the anharmonicity of the oscillations is also a distinct feature of the nonlinear large amplitude regime of the coalescence. The anharmonicity is clearly seen only in the nonlinear regime. The theory of \citet{1987ApJ...321.1031T} is constructed under the assumption that the gas pressure can be ignored during the implosion, but the gas pressure significantly increases at the center of the current sheet and produces two MHD fast-mode shocks \citep{1982JPlPh..27..491F,2015ApJ...807..159T}. 

{Now numerical modelling of plasmoid coalescence is an essential method to advance our understanding of particle acceleration associated with plasmoid dynamics. \citet{2006Natur.443..553D} pointed out that contracting plasmoids can be an efficient accelerator via Fermi acceleration. \citet{2016ApJ...820...60G} applied the idea to plasmoid-dominated reconnection in a simulated eruptive flare using a 2.5D MHD simulation, and found that energy gain of electrons in plasmoids can be higher than the previous estimation by \citet{2006Natur.443..553D} due to strong plasma compression that occurs at the flare current sheet. However, \citet{2010ApJ...714..915O} performed 2D particle-in-cell simulations of magnetic reconnection with multiple plasmoids and found that the key process for electron acceleration is the secondary magnetic reconnection at merging points formed between coalescing plasmoids. The idea around the interaction between plasmoids and shocks by \citet{2013PhRvL.110e1101N} has not been examined numerically. Therefore, our knowledge regarding electron acceleration has not been established yet. More detailed numerical modelling and comparison with observations should be important for understanding the origin of QPPs in non-thermal emissions.}

\subsection{Equivalent LCR contour} \label{sec : LCR mechanism}

An alternative, non-hydrodynamic mechanism for QPPs, based upon the consideration of a flaring magnetic configuration as an equivalent LCR contour  was proposed by \citet{1998A&A...337..887Z}. The model is based on the difference in the values of the electrical conductivity in the coronal and chromospheric parts of the flaring region. In the corona, the parallel conductivity is much higher than across the field, and the electric current should go along the magnetic field lines. In the partly-ionised photospheric plasma, the electric current could go across the field. Hence there appears a closed electrical circuit formed in a coronal loop by a field-aligned current, which goes from one footpoint to the other, and the cross-field current between the footpoints in the photosphere. 

Dynamics of the electric current ${\cal I}$ in such a contour is described by the equation:
\begin{equation}
\label{LCRode}
\frac{1}{c^2} {\cal L} \frac{d^2 {\cal I}}{d t^2} + {\cal R}({\cal I}) \frac{d{\cal I}}{d t} + \frac{1}{{\cal C}({\cal I})} {\cal I} = 0,
\end{equation}
where $c$ is the speed of light, 
\begin{eqnarray*}
\frac{1}{{\cal C}} =  \frac{{\cal I}^2 l_\mathrm{c}^2}{\pi c^4 \rho_0 r_\mathrm{c}^4}\Big(1 + \frac{c^2 r_\mathrm{c}^2 B_{z0}^2}{4 {\cal I}^2}\Big),\quad {\cal L} = 2l_\mathrm{c}\Lambda,
% \mathrm{\ \ \ \ } \Lambda =  \mathrm{ln}\frac{4l_\mathrm{c}}{\pi r_\mathrm{c}} - \frac{7}{4},
\end{eqnarray*}
are the effective circuit capacitance and inductance, respectively; {{where}}  $B_{z0}$ is the magnetic field along the axis of the loop; $r_\mathrm{c}$ and $l_\mathrm{c}$ are the minor radius and length of the coronal part of the loop, respectively;  $\rho_0$ is the mass density in the coronal part of the loop \citep{2009SSRv..149...83K} {{and  $\Lambda =  \mathrm{ln}\frac{4l_\mathrm{c}}{\pi r_\mathrm{c}} - \frac{7}{4}$.}} The nonlinear term ${\cal R}({\cal I})$ represents the effective resistance. It combines the resistance connected with the ion-neutral collisions in the photosphere, proportional to ${\cal I}^2$, and the electromotive force associated with the photospheric convection. For sufficiently large photospheric convective flows the expression for ${\cal R}({\cal I})$ can become negative, inducing the alternate current in the system. This effect is more pronounced in the loops with a lower equilibrium electric current.

%\begin{eqnarray}
%\frac{1}{{\cal C}} =  \frac{{\cal I}^2 l_\mathrm{c}^2}{\pi c^4 \rho_0 r_\mathrm{c}^4}\Big(1 + \frac{c^2 r_\mathrm{c}^2 B_{z0}^2}{4 {\cal I}^2}\Big),\quad \\
%{\cal L} = 2l_\mathrm{c}\Lambda, \mathrm{\ \ \ \ } \Lambda =  \mathrm{ln}\frac{4l_\mathrm{c}}{\pi r_\mathrm{c}} - \frac{7}{4},
%\end{eqnarray}

In the linear regime, when the effective resistance ${\cal R}$ and capacitance ${\cal C}$ are independent of the amplitude of the alternate current, equation (\ref{LCRode}) is a damped harmonic oscillator equation. It describes an alternate electric current with the period:
\begin{equation}
\label{pLCR}
P_\mathrm{LCR} \approx \begin{cases}
\displaystyle\frac{ 4\pi r_\mathrm{c} \sqrt{2\pi \Lambda  \rho_\mathrm{c}}}{B_{z0}},  \mbox{\ \ in an untwisted loop,} \\ 
\mbox{}\\
\displaystyle\frac{2\pi r_\mathrm{c}^2 \sqrt{2\pi \Lambda \rho_\mathrm{c}}}{{\cal I}_{0}}, \mbox{\ \ in a highly twisted loop,}
	\end{cases}
\end{equation}
where ${\cal I}_0$ is the equilibrium electric current in the loop, 
which is connected with the azimuthal component of the equilibrium magnetic field. For different values of the parameters, the model gives periods from a fraction of a second to a few minutes \citep[e.g.][]{2008PhyU...51.1123Z}. 

The LCR oscillations of a flaring magnetic loop can produce QPPs of both thermal and non-thermal emission. In particular, the intensity of the microwave emission produced by the gyrosynchrotron mechanism, is proportional to the angle between the line-of-sight and the local magnetic field. Thus, periodic variations of the field-aligned electric current that can be considered as the periodic appearance of the azimuthal component of the magnetic field, should result in the modulation of the gyrosynchrotron emission. The Ohmic heating of the alternate current would lead to a periodic variation of the plasma temperature. The temperature variation period is two times shorter than the period of the electric current oscillations, as it is proportional to the current amplitude squared. The associated variation of the electric field may result in periodic acceleration of charged particles, provided the electric field exceeds the Dreicer field. The periodic acceleration will lead to the periodic non-thermal emission. Another possibility is the transverse perturbation of the axis of the loop, caused by the periodic twisting of the loop by the alternate electric current. Indeed, twisting leads to the deformation of the loop's plane, i.e. making the loop having an S-shape. Hence, periodic alternate twisting should periodically deform the loop's plane, resembling a kink oscillation, with the structure resembling the second or higher harmonics. The interaction of this oscillation with a magnetic X-point situated nearby, could lead to the periodic modulation of the reconnection rate.

This mechanism may also easily explain drifts of the QPP periods, that are often observed in flares. For example,  \citet{2017ApJ...836...84D} detected  high-quality oscillations of the time derivative of the soft X-ray emission of the decay phase of a flare, with the period gradually increasing from 25~s to 100~s. If the detected QPPs are produced by the alternate current, a gradual decrease in the equilibrium current ${\cal I}_0$, caused, for example, by Ohmic dissipation, would lead to the increase in the oscillation period, see Eq.~(\ref{pLCR}).

Similar ideas could be applied to the oscillatory interaction of two or several current-carrying loops, via the variation of the mutual inductance \citep[e.g.][]{2005A&A...433..691K, 2009SSRv..149...83K}. This interaction would result in variation of the electric currents, and also of the geometrical parameters of the magnetic configuration, for example the variation of the loop plane, the distance between individual loops, etc., which would look like transverse oscillations. It was shown that the inductive interaction inside a flaring active region may lead to longer period oscillations, of several minutes. The observational manifestation of these oscillations could, in particular, lead to QPPs in flaring light curves, either directly, by the effect of the alternate electric current, or indirectly, by, for example, periodically induced magnetic reconnection. 

As was emphasised by \citet{2009SSRv..149...83K}, the equivalent LCR circuit approach ignores the fact that changes of the magnetic field and related electric current propagate as torsional Alfv\'en waves at the Alfv\'en speed, assuming the instant changes of the electric current in the whole circuit. Therefore, this model describes adequately the oscillations and evolution with time scales longer than the Alfv\'en travel time along the loop. 

\subsection{Autowave processes in flares}\label{sec : autowave}

The effect of autowaves has not been directly addressed in the context of solar flares so far. However, the possibility of the triggering of flaring energy release by an MHD wave, and the subsequent excitation of the wave by the flare, creates a theoretical ground for the occurrence of autowave processes.

One possible manifestation of such a process is connected with the well-established progression of the flaring hard X-ray and EUV brightenings along the photospheric neutral line in two-ribbon flares. The typical speed of this {{progression along the ribbons}} is a few tens of km/s {{(e.g. \citealt{2003ApJ...595L.103K}, \citeyear{2005AdSpR..35.1707K}; \citealt{2005ApJ...630..561B}; \citealt{2005ApJ...625L.143G}; \citealt{2009ApJ...693..132Y}; \citealt{2013ApJ...777...30I})\footnote{Note that this propagation along the ribbons is a different phenomenon to the  gradual separation motion of the flare ribbons, which is (also) typically of the order of tens of km/s.}}}. \citet{2011ApJ...730L..27N} suggested that this value of the speed is well consistent with the perpendicular group speed of a highly oblique slow magnetoacoustic wave. It was found  that the perpendicular group speed has a rather sharp maximum reaching about 10--20\% of the sound speed for the propagation angles of about 25--28~degrees to the magnetic field. In a typical flaring plasma of temperature 10$^7$~K and with Alfv\'en speed of 1000~km/s, the highest perpendicular group speed of a slow magnetoacoustic wave is about 40~km/s.

In the proposed scenario a primary energy release occurring somewhere above the neutral line, in the flaring arcade, excites a slow magnetoacoustic pulse that propagates downward. The pulse gets reflected from the chromosphere and returns back to the top of the flaring arcade, slightly offset from the location of the primary energy release along the neutral line \citep{2011A&A...536A..68G}. There the wave triggers another energy release by one of the mechanisms described in \S\ref{periodic triggering}. This energy release excites another slow wave, causing the next cycle of this autowave mechanism. Along the field, the pulse propagates at a speed close to the sound speed, while across the field its group speed is 10--20\%  which is consistent with the observed speed of the brightening progression along the neutral line. In addition, this mechanism explains readily the quasi-periodic nature of the energy releases. The oscillation period would be determined by the acoustic travel time from the footpoints to the arcade top, e.g. 30--80~s for typical solar flares. This value is consistent with the quasi-periodic progression of hard X-ray sources along the neutral line, observed by \citet{2005ApJ...625L.143G}. An important element of this mechanism is that, in contrast with the fast magnetoacoustic waves propagating in a non-uniform plasma (see \S\ref{wavetrain}), slow waves experience very weak dispersion, and hence the initial pulse does not evolve in a quasi-periodic wave train. 

This mechanism could be modified. For example, the excitation of the slow magnetoacoustic pulse could occur not near the energy release site, but at the chromosphere, by the precipitating non-thermal particles accelerated by the energy release near the top of the arcade. Some asymmetry of the footpoints with respect to the arcade top would cause some difference in the acoustic travel time in the opposite magnetic legs. It would produce the double-peak structure of the emission peaks in the flare light curve, which is a frequently detected feature of QPPs. 

\section{Conclusions}
      %include the cartoon (needs updating). 
     %%% CARTOON ? %%%%%%%%%%
 
%perhaps summarise/review these?  => TABLE ??????
% When describing/detailing each mechanism, I think it would be good to state clearly:
      % * What range of periods the mechanism can create/estimate/address.\subsubsection{What range of periods can the mechanism generate?}
      % * What underlying physics determines these periodicities
      % * How much this mechanism can be proven/identified in observations.
      % * What theory/advances should be done in the modelling as a next step, e.g. parametric studies.
 
% Conclusive proof will require identification of multiple characteristics in a single observed event with a favourable magnetic configuration. 

There are quasi-periodic patterns in solar and stellar flaring energy releases. Often the EM radiation generated in flares shows a pronounced oscillatory pattern, with characteristic periods ranging from a fraction of a second to several minutes, or even longer in the case of stellar superflares. These are referred to as quasi-periodic pulsations (QPPs) to emphasise that they often contain apparent amplitude and period modulation. QPPs have been detected in all EM bands (including radio, microwave, white-light, H$\alpha$, UV, EUV, soft X-ray, hard X-ray and gamma-ray) and occur in all stages of the flare. QPPs are detected across multiple instruments, so they are not an instrumental effect, and are observed in a significant fraction of flares.  This review paper is primarily a theoretical {\emph{modelling}} review and details the possible physical mechanisms underpinning QPPs, with an emphasis on the physical processes that generate the resultant range of periodicities. 

We have reviewed eleven potential physical mechanisms underpinning QPP generation (\S\ref{physmechqpp}). These can be classed according to the nature of the underlying physical process:
\begin{itemize}
\item{
{\emph{Oscillatory}} processes of the emitting plasma, including  MHD oscillations (\S\ref{sec:coronal_seismology}), QPPs triggered periodically by external waves (\S\ref{periodic triggering}), dispersive wave trains (\S\ref{wavetrain}), the magnetic tuning fork (\S\ref{sec : Magnetic tuning fork})   and the equivalent LCR contour (\S\ref{sec : LCR mechanism}). Oscillations are (quasi-)periodic motions around an equilibrium, connected with the competition between inertia and an effective restoring force. Properties of oscillations (spectrum, amplitudes, phases, etc) are prescribed by the initial perturbation. The advantage of the MHD oscillation explanation  is the observation of multiple periodicities.
}
\item{
{\emph{Self-oscillatory}} processes, including  the \lq\lq{load-unload}\rq\rq{} model and relaxation processes. This includes periodic or repetitive spontaneous reconnection, including oscillatory reconnection (\S\ref{sec:OscillatoryReconnection}), thermal overstabilities (\S\ref{theros}) and wave-flow overstabilities (\S\ref{waveflow}),  wave-driven reconnection in the Taylor problem (\S\ref{sec : wave-driven reconnection in the Taylor problem}) as well as the coalescence of two magnetic flux tubes (\S\ref{sec : Two loop coalescence}). Mathematically, self-oscillations are associated with a limit cycle. Usually self-oscillations have properties that are independent of the initial excitation and occur in essentially non-conservative systems. The self-oscillation period may depend upon the amplitude. In flares, a steady inflow of magnetic flux towards a reconnection site could result in repetitive magnetic reconnection (\lq\lq{magnetic dripping}\rq\rq{}) that should be considered a self-oscillatory process. The energy supply for self-oscillations comes from an essentially non-periodic source. In  the \lq\lq{load-unload}\rq\rq{} model, QPPs are a side-effect of the transient energy release, connected with the relationship of the energy load-and-then-unload balance. The advantage of the time-dependent reconnection model  is the natural explanation of the simultaneity of QPPs in different bands (as they are produced by the same cause: the time-varying rate of the electron acceleration).
} 
\item{
We also considered {\emph{autowave}} processes in flares (\S\ref{sec : autowave}). Properties of autowaves are independent or weakly-dependent on the initial excitation, and so they are determined only by the parameters of the system.
}
\end{itemize}
Terminology and definitions were given in \S\ref{defs}.

\subsection{Future directions and key unanswered questions}

There remain key unanswered questions concerning QPPs, including:
\begin{itemize}
\item{
Is there any statistical relationship between QPP parameters (periods, decay times, relative and absolute amplitudes, modulations, etc) with the parameters of the host flare?
}
\item{
The observed periods of QPPs coincide by the order of magnitude with the MHD oscillations and waves detected abundantly in the solar corona (and well resolved in time and space). These MHD oscillations and waves typically have a few percent relative amplitude. In contrast, QPPs can reach a modulation depth of up to 100\%. If QPP are caused by MHD oscillations, how can the oscillatory signal be amplified?
}
\item{
Are QPPs detected in different phases of the flare fundamentally different? Are QPPs detected in thermal and non-thermal emission different?
}
\item{
Can we distinguish between different classes of QPPs (if indeed there are different classes)?
}
\item{
Are the QPPs detected in stellar superflares, that are much more powerful than the strongest detected solar flares, produced by the same mechanisms as in solar flares?
}
\end{itemize}

The occurrence of QPPs puts additional constraints on the interpretation and understanding of the fundamental processes operating in both solar and stellar flares, e.g. particle acceleration and magnetic energy liberation. Simply put, there must be a physical reason for the flaring emission being arranged in a sequence of quasi-periodic bursts. The importance of a full understanding of QPPs is essential in order to work towards an integrated model of solar and stellar flares, as well as unlocking  a potential diagnostic of the flare process.

When reviewing the QPP physical mechanisms in this paper, we have emphasised (where possible) the following details: [i] What range of periods can the mechanism generate? [ii] What underlying physics determines these periodicities? [iii] How much can this mechanism be proven/identified in observations? All the QPP physical mechanisms detailed in \S\ref{physmechqpp} require further study and refinement, e.g. parametric studies and forward modelling of produced observables. There is currently no  physical mechanism that can unambiguously explain all QPPs, and conclusive proof will require identification of multiple characteristics in a single observed event with a favourable magnetic configuration. In this context, the main advantage of solar observations, the availability of spatial information about the plasma and magnetic structures in the flaring region, and also of the sources of different emissions, opens up very interesting perspectives and needs full exploitation. Another important feature of QPP, the non-stationarity of the period and amplitude of the oscillatory patterns, in other words, the \lq\lq quasi\rq\rq-ness, requires the development of new analytical techniques addressing the intrinsically non-stationary nature of QPPs. 

%-------------------------------------------------------------------------------
 \begin{acknowledgements}
This review arose from discussions at a workshop on \lq{Integrated Plasma Modelling of Solar Flares}\rq{} at the Lorentz Center, University of Leiden (May 2015). JAM acknowledges generous support from the  Leverhulme Trust and this work was funded by a Leverhulme Trust Research Project Grant: RPG-2015-075. {{JAM also acknowledges this material is based upon work supported by the US Air Force Office of Scientific Research, Air Force Material Command, USAF under Award No. FA9550-16-1-0032.}} JAM acknowledges IDL support provided by STFC. VMN acknowledges the support from the European Research Council under the SeismoSun  Research Project No. 321141, and the BK21 plus program through the National Research Foundation funded by the Ministry of Education of Korea. {{This work was supported in part by the Russian Foundation for Basic Research grant No. 17-52-80064 (VMN).}}  MD acknowledges the IAP P7/08 CHARM programme as well as the SIDC Data exploitation programma from PRODEX. PJ acknowledges support from grant 16-13277S of the Grant Agency of the Czech Republic. JAM and VMN acknowledge the support of ISSI. ST acknowledges support by the Research Fellowship of the Japan Society for the Promotion of Science (JSPS).

\end{acknowledgements}

\bibliographystyle{aps-nameyear}

%\bibliography{Lorentz_waves_flares_bibliography_11may2017}

\bibliography{bibliography_QPPs_review}

\appendix

\section{Global waves generated by flares: shock waves, blast waves and \lq{flare waves}\rq}\label{Appendix : global_flare_waves} 

On the global scale, we can consider flares to  be enormous impulsive energy releases in an elastic and compressive medium surrounding the flaring site. Simply put, we expect MHD waves and shocks to propagate away from the flaring region. \citet{2008SoPh..253..215V} provide an excellent review of the origin of large-scale coronal shock waves, and detail the physical mechanisms capable of launching MHD shocks.  Of relevance, \citet{2008SoPh..253..215V} review the idea of a shock wave driven by a 3D piston effect, where the expanding driver pushes the plasma in all directions (piston-shock). As a special case, when the driver is of finite duration (temporary piston)  this generates a freely-propagating simple-wave shock, also known as a {\emph{blast wave}}. Thus, a flare modelled as a (explosion-like) pressure pulse would generate a blast wave or shocked simple-wave (i.e. the driver has an acceleration phase, deceleration phase and then stops). Whereas a CME-driven shock is a piston-shock in its early stages (and in later stages, a piston-shock in combination with a bow shock). The physical difference is that the piston-driven shock wave  is drawing additional energy from the piston (source region) whereas in a blast wave the shock is freely propagating and there is no additional energy input (see also \citealt{1959flme.book.....L}).

Global-scale propagating disturbances have been observed directly in the corona initially with the SoHO/EIT instrument and became known as {\it{EIT waves}}. These are bright, wave-like (pulse) features propagating globally across the solar disk through the corona (\citealt{1997SoPh..175..571M}; \citealt{1998GeoRL..25.2465T}). After being observed by different instruments, EIT waves later became known as large-scale Coronal Bright Fronts (CBFs, see reviews by \citealt{2011SSRv..158..365G}; \citealt{2015LRSP...12....3W}) and also as  Global Coronal Waves (\citealt{1999SoPh..190...91H}; \citealt{2016GMS...216..381C}).

In addition to EIT waves, there are {\it{Moreton waves}} (\citealt{1960AJ.....65U.494M}; \citealt{1960PASP...72..357M}). These are {{ propagating, bright fronts in H{$\alpha$} line center and blue wing (and dark fronts in H{$\alpha$} red wing)}} and, given that the H{$\alpha$} spectral line is formed in the chromosphere,  are a chromospheric phenomenon. The study of EIT waves and their association, or not, with chromospheric Moreton waves is a subject of active research (e.g. \citealt{2011A&A...531A..42L}; \citealt{2013SoPh..288..567L}; \citealt{2014SoPh..289.3279L}) and readers are referred to a comparison of different EIT wave models (\citealt{2017SoPh..292....7L}). What seems to be clear is that  Moreton waves are related to CMEs (\citealt{2002ApJ...572L..99C}; \citealt{2016GMS...216..381C}). But what about the link to flares, specifically? Since Moreton waves were discovered before CMEs were discovered  \citep{1973spre.conf..713T} solar flares were initially thought to be the cause of Moreton waves \citep{1966AJ.....71..197R}. \citet{1968SoPh....4...30U} proposed that the pressure pulse in the solar flare generates a fast wave propagating in the corona. As the wavefront sweeps through the chromosphere, it pushes  chromospheric material downward and this is how a Moreton wavefront is formed. Thus, Moreton waves were thought to be {{blast waves}} (freely-propagating shock wave). Moreover, chromospheric Moreton waves were given the specific terminology  {\emph{flare waves}} (e.g. \citealt{1967SoPh....1...66Z}; \citealt{2001ApJ...560L.105W}; \citealt{2004A&A...418.1101W}; \citealt{2004A&A...418.1117W}). Note that under this interpretation, the chromospheric Moreton wave is the footprint of the fast-mode EIT wave. This coronal counterpart (i.e. the fast-mode EIT wave) then also took on the terminology {\emph{coronal flare wave}} and {\emph{coronal Moreton wave}} (e.g. \citealt{2000SoPh..193..161T}; \citealt{2002A&A...394..299V}).

However, this interpretation has now been superseded by the link to CMEs, rather than driven by flares. E.g. \citet{2002ApJ...572L..99C} {{replace}} the  blast wave (initiated by the solar-flare pressure pulse) by a piston-driven shock wave from a CME (see \citealt{2008SoPh..253..215V} and \S2.2 of \citealt{2015LRSP...12....3W} for further details). Furthermore, \citet{2006ApJ...641L.153C}  selected 14 M-class and X-class flares that were not associated with CMEs and found that none of the flares was associated with any EIT waves. \citeauthor{2006ApJ...641L.153C} {{et al. conclude that it is unlikely that pressure pulses from flares generate EIT waves.}} Thus, even though the current evidence favours CMEs as the origin of EIT waves rather than flares, there is still the possibility that some waves are generated by flare-associated pressure pulses (e.g. \citealt{2012ApJ...753...52L}; \citealt{2013SoPh..288..255K}; and see \S5.1 of \citealt{2015LRSP...12....3W}). It is seen that there are many types of waves in the corona, some are driven by CMEs, and some are by flares. Even for CME-associated waves, it seems that {{there are two distinct types of EUV coronal global waves,}}  where the faster  EUV global waves are interpreted as  nonlinear fast waves (driven by the impulsive expansion of an erupting CME) and {{that}} there also exists a slower type of EUV wave (see reviews by \citealt{2015LRSP...12....3W}; \citealt{2016GMS...216..381C}; \citealt{2017SoPh..292....7L}, and references therein).

% Recently, it has been claimed that {{there are two distinct types of EUV coronal global waves,}}  where the faster  EUV global waves are interpreted as  nonlinear fast waves (driven by the impulsive expansion of an erupting CME) and {{that}} there also exists a slower type of EUV wave (see reviews by \citealt{2015LRSP...12....3W}; \citealt{2016GMS...216..381C}; \citealt{2017SoPh..292....7L}, and references therein).

%"Recently, it has been claimed that" with "It is seen that there are many types of waves in the corona, some are driven by CMEs, and some are by flares. Even for CME-associated waves, it seems that ...".

%Recently, it has been claimed that EIT waves may instead be two distinct types of EUV coronal global waves;  where the faster  EUV global waves are interpreted as  nonlinear fast waves (driven by the impulsive expansion of an erupting CME) and there also exists a slower type of EUV wave (see reviews by \citealt{2015LRSP...12....3W}; \citealt{2016GMS...216..381C}; \citealt{2017SoPh..292....7L}, and references therein).

%\citeauthor{2006ApJ...641L.153C} {{et al.}} concludes that it is unlikely that pressure pulses from flares generate EIT waves, thus strengthening the interpretation that EIT waves are piston-shocks associated with CMEs, not blast waves associated with flares.

\section{Global, flare-generated waves in the solar interior: sunquakes}\label{Appendix : sunquakes}

Another global wave-like phenomenon associated with flares is that of {\emph{sunquakes}}. \citet{1972ApJ...176..833W} proposed that solar flares could excite free global oscillations inside the Sun similar to earthquakes, i.e. that flares should deliver acoustic impulses to the solar interior  \citep{2011SSRv..158....5H}. These  seismic transients were modelled by \citet{1995ESASP.376b.341K} and then  were first discovered by \citet{1998Natur.393..317K}. Sunquakes are seismic (acoustic)  waves  generated  by flares and  manifest  themselves  at  the photospheric solar  surface,  as  a  quasi-circular pattern of ripples moving away from the flare epicentre. \citet{2011SSRv..158..451D} reports that the manifestation of this acoustic energy is seen 20--50 Mm from the source when it refracts back to the solar surface within about an hour after the commencement of the flare. This refraction is a result of the increasing sound speed with increasing depth in the solar interior. Thus, flare-driven sunquakes open the prospect of using seismology (helioseismic analysis) to study the solar interior structure \citep{2008SoPh..251..627L} as well as informing the general topic of MHD wave behaviour in inhomogeneous media. See \citet{2011LNP...832....3K} for a comprehensive review of the basic principles of global and local helioseismology and see  \citet{2006SoPh..238....1K} for a review of the properties of sunquakes. For  examples of sunquakes see, e.g., \citet{2008SoPh..251..613M} and examples by \citet{2014ApJ...796...85J}, \citet{2015ApJ...812...35M}, and \citet{2015SoPh..290.3151B}.

\citet{2014arXiv1402.1249K} reports that the excitation impact strongly correlates with the impulsive flare phase and is caused by the energy/momentum transported from the energy-release site(s) but that the physical mechanism is currently uncertain. These  seismic transients have been explained via a \lq{thick-target}\rq{ } hydrodynamic model (see e.g. \citealt{1975SvA....18..590K}; \citealt{1981SoPh...73..269L}; \citealt{1985ApJ...289..425F}). Here, a  beam of high-energy particles is accelerated in the corona, heats the chromosphere, which results in a compression of the lower chromosphere. This compression produces chromospheric evaporation and a downward-propagating shock wave (velocity impulse) which impacts the photosphere (i.e. a hydrodynamic impact)  causing the seismic response (\citealt{1998Natur.393..317K}; \citealt{2014arXiv1402.1249K}; \citealt{2015SoPh..290.3163Z}).

\citet{2011SSRv..158..451D} details several alternative generation mechanisms for sunquakes that have been proposed, including a generation mechanism based on the direct interaction of high-energy particles (electrons or protons) with the photosphere (\citealt{2005ApJ...630.1168D};  \citealt{2007ApJ...664..573Z});  pressure transients related to photospheric backwarming  by enhanced chromospheric radiation  (\citealt{2000SoPh..192..261L}; \citealt{2005ApJ...630.1168D}); flare acoustic emission due to  impulsive heating of the low photosphere and radiative backwarming  \citep{2006SoPh..239..113D}; and a magnetic jerk that manifests as a seismic response occurring  during the re-organisation of the magnetic topology, specifically a change in field line inclination at the footpoints  (\citealt{2008ASPC..383..221H}, and see  recent extension by \citealt{2016ApJ...831...42R}).

\citet{2011SSRv..158..451D} reports that (with current instruments) sunquakes  are a rare phenomenon and most flares do not generate detectable seismic emission in the p-mode spectrum. However, \citet{2014arXiv1402.1249K} speculates that perhaps all flares generate some seismic response, but if the amplitude is not high enough the signal may be lost in the background noise. \lq{Starquakes}\rq{ } resulting from stellar flares have also been detected on other stars \citep{2014IAUS..301..349K}, while other studies are less optimistic \citep[e.g.][] {2015MNRAS.450..956B}. If detected confidently, starquakes could provide new asteroseismic information and impose rigorous constraints on stellar flare mechanisms.
%----------------------------------------

\end{document}